\documentclass[12pt]{amsart}
\usepackage{amsmath}
\usepackage{amssymb}
\usepackage{amsthm}
\usepackage[backend=biber]{biblatex}
\usepackage{hyperref}
\usepackage{mathtools}
\usepackage{mathrsfs}
\theoremstyle{plain}
\newtheorem{theorem}{Theorem}
\newtheorem{lemma}[theorem]{Lemma}
\newtheorem{proposition}[theorem]{Proposition}

\theoremstyle{definition}

\theoremstyle{remark}

\allowdisplaybreaks

\title{Analysis of sum-of-squares relaxations for the quantum rotor model}
\author{Sujit Rao}
\address{77 Massachusetts Avenue\\Cambridge MA, 02139}
\email{sujit@mit.edu}
\begin{document}

\begin{abstract}
The noncommutative sum-of-squares (ncSoS) hierarchy was introduced by Navascu\'{e}s--Pironio--Ac\'{i}n as a sequence of semidefinite programming relaxations for approximating values of noncommutative polynomial optimization problems, which were originally intended to generalize quantum values of nonlocal games. Recent work has started to analyze the hierarchy for approximating ground energies of local Hamiltonians, initially through rounding algorithms which output product states for degree-2 ncSoS applied to \textsc{Quantum Max-Cut}. Some rounding methods are known which output entangled states, but they use degree-4 ncSoS. Based on this, Hwang--Neeman--Parekh--Thompson--Wright conjectured that degree-2 ncSoS cannot beat product state approximations for \textsc{Quantum Max-Cut} and gave a partial proof relying on a conjectural generalization of Borrell's inequality. In this work we consider a family of Hamiltonians (called the quantum rotor model in condensed matter literature or lattice $O(k)$ vector model in quantum field theory) with infinite-dimensional local Hilbert space $L^{2}(S^{k - 1})$, and show that a degree-2 ncSoS relaxation approximates the ground state energy better than any product state.
\end{abstract}

\maketitle


\vspace{-2em}
\section{Introduction}
\subsection{Classical polynomial optimization}
Low-degree polynomial optimization problems (POPs) are ubiquitous in both theory and practice but are NP-hard, motivating the construction of algorithms which can efficiently output approximate solutions. One of the most common approaches is to use the sum-of-squares (SoS) hierarchy, which is a sequence of semidefinite programs whose size is polynomial (so can be solved efficiently using standard convex optimization methods) and whose solutions approximate the optimum of a POP. This hierarchy is best studied for constraint satisfaction problems (CSPs), which have a canonical form as a low-degree POP over the boolean hypercube $\{\pm 1\}^{n}$. For many CSPs it is known that the first SDP in the SoS hierarchy gives the optimal algorithm under the unique games conjecture, a common complexity-theoretic assumptions. 

POPs over continuous domains are perhaps even more common than POPs over the boolean hypercube. For example, POPs over the sphere have several applications in quantum information and complexity, and have been studied from both quantum and classical perspectives. However, less is known about how well the SoS hierarchy performs over continuous domains as compared to the nearly-complete results for CSPs. Interestingly, for the ``rank-$k$ Grothendieck problem'' over a product of spheres, the best known results \cite{Briet_Oliveira_Vallentin_2014} are based on adapting techniques for classical CSPs and \textsc{Max-Cut} \cite{Goemans_Williamson_1995} in particular.

\subsection{Noncommutative polynomial optimization}
Many problems in quantum information can be recast as ``noncommutative POPs,'' which were first defined in \cite{Pironio_Navascues_Acin_2010} and for which there is a similar ``noncommutative SoS'' (ncSoS) hierarchy of SDPs to approximate their values. Initial work focused on ncPOPs arising from quantum nonlocal games, motivating the recent proof of $\mathsf{MIP}^{*} = \mathsf{RE}$ in quantum complexity \cite{Ji_Natarajan_Vidick_Wright_Yuen_2020} and implying that the hierarchy may not always converge to the intended value -- in this case the tensor product model is considered to be more physical and one may be more interested in computing its value than the commuting operator value of a nonlocal game, but $\mathsf{MIP}^{*} = \mathsf{RE}$ implies that no algorithm, including ncSoS, can compute the tensor product value. In particular this also implies that no algorithm can compute the value of a ncPOP where the variables are restricted to be finite-dimensional matrices, since such an algorithm could compute the tensor product value of a nonlocal game.

More recent work has begun to analyze the hierarchy when applied to POPs arising from local Hamiltonians, which are usually thought of as canonical quantum analogs of CSPs. The earliest result in this area is \cite{Brandao_Harrow_2016}, which adapted a classical correlation-reducing technique to construct product states from an SDP solution. Later work \cite{Gharibian_Parekh_2019} used \cite{Briet_Oliveira_Vallentin_2014} to more directly construct product-states for \textsc{Quantum Max-Cut} (equivalent to the quantum Heisenberg model with a change of constants) from the first level of ncSoS.
Further work \cite{Anshu_Gosset_Morenz_2020, Parekh_Thompson_2021, Parekh_Thompson_2022, King_2022, Lee_2022} has mainly used the second or higher levels of the ncSoS hierarchy to feed a product state into a low-depth entangling circuit. It is widely believed that higher levels are necessary to construct good entangled states, at least for \textsc{Quantum Max-Cut} \cite{Hwang_Neeman_Parekh_Thompson_Wright_2023}, and a simple integrality gap using star graphs is also known, with using the graph as ``hard instance'' having first been discussed in \cite{Anshu_Gosset_Morenz_2020} and the calculation of the integrality gap first appearing in print in \cite{Parekh_Thompson_2022}.

\subsubsection{Results} This work introduces a new technique, inspired by Connes's noncommutative geometry, for constructing entangled states directly, in a single step, from the output of an ncSoS SDP. This is a noncommutative analog of the technique used in \cite{Briet_Oliveira_Vallentin_2014, Goemans_Williamson_1995} for the rank-$k$ Grothendieck problem and classical \textsc{Max-Cut}. We use this technique to show that the first ncSoS level approximates the ground state energy better than any product state for the quantum rotor model from condensed matter physics (also called the lattice $O(n)$ nonlinear sigma model and lattice $O(n)$ vector model in quantum field theory). This stands in stark contrast to \textsc{Quantum Max-Cut}, where the star graph integrality gap for level-1 ncSoS matches the product state gap.

In the rest of this introduction, we describe the model and state the results and motivation more precisely.

\subsection{Motivation for continuous-variable Hamiltonians}
Local Hamiltonians are usually taken to be finite-dimensional in order to simplify the applications to complexity, even though the majority of Hamiltonians relevant to practical applications and quantum simulation are infinite-dimensional (since the underlying system has continuous-variable degrees of freedom) and are more analogous to classical POPs over continuous domains. 
For example, \textsc{Quantum Max-Cut} is based on the quantum Heisenberg model of ferromagnetism, but the exchange interaction in the Heisenberg model is in fact a perturbative approximation to the continuous-variable kinetic energy term and Coulomb interaction under the assumption that electrons have fermionic statistics and are bound to their nuclei \cites[Chapter 2]{Mattis_1981}[Chapter 9]{Du_Tremolet_de_Lacheisserie_Gignoux_Schlenker_2002}[Section 3.2.1]{Diep_2013}. Moreover, some ferromagnetic materials such as iron or ferric oxide are conductors, so valence electrons are not bound to their nuclei and the lattice approximation to continuous space in the Heisenberg model becomes invalid. More accurate calculations of the Curie temperature or magnetic moment of ferromagnetic materials usually use the Stoner model \cite{Stoner_1938}, which does not correspond to a Hamiltonian.

Ultimately all quantum models for physically relevant systems can be derived from quantum field theory with continuous degrees of freedom. While finite-dimensional approximations can exist and are similar to classical CSPs, the original infinite-dimensional model may be more accurate and it is certainly not an issue to solve the original infinite-dimensional problem if doable. As the work in this paper on the quantum rotor model shows, it may potentially be beneficial to instead solve infinite-dimensional problems, and the ncSoS hierarchy applied to them may be more practically relevant than previously thought.

\subsubsection{Prior work on complexity of infinite-dimensional or continuous-variable Hamiltonians}
There has been earlier work studying continuous-variable Hamiltonians with infinite-dimensional local Hilbert space from complexity-theoretic or computer science perspectives. For example, the hardness of the Bose-Hubbard model was studied in \cite{Childs_Gosset_Webb_2014}. The Bose-Hubbard Hamiltonian has local Hilbert space $L^{2}(\mathbb{R})$, usually thought of as $L^{2}(\mathbb{N})$ with the isomorphism given by the eigenstates of the quantum Harmonic oscillator, and has the form
\[ b\sum _{(v, w) \in E} a_{v}^{*}a_{w} + c \sum _{v \in V} n_{v}(n_{v} - 1) \]
where $a_{v}$ is the creation operator on site $v$ and $n_{v}$ is the number operator. The Hamiltonian commutes with the total number operator $\sum _{v} n_{v}$, so it can be block-diagonalized into finite-dimensional blocks corresponding to total particle number. The $\mathsf{QMA}$-hardness result shown in \cite{Childs_Gosset_Webb_2014} includes the total particle number as a parameter, which is one option for converting the original continuous-variable problem into a finite-dimensional problem. However, it would be desirable to have a framework for showing $\mathsf{QMA}$-hardness of continuous-variable Hamiltonians parameterized by a discrete input (such as a graph for the Bose-Hubbard or quantum rotor models).

This Hamiltonian has some similarities to the quantum rotor Hamiltonian: it involves a 2-local interaction term summed over edges of a graph and a 1-local term summed over each vertex, with nonnegative weights for both sums. While the quatum rotor model is most likely in $\mathsf{StoqMA}$ and the Bose-Hubbard model is $\mathsf{QMA}$-hard, it would still be interesting to study algorithms for solving either one, or other infinite-dimensional Hamiltonians.

\subsubsection{Classical motivation for continuous-variable Hamiltonians}
There are also purely classical problems which motivate the consideration of continuous-variable Hamiltonians, naturally arising in classical problems with continuous degrees of freedom. The quantum rotor model has a kinetic term, which can be thought of as a perturbation to the objective function of the rank-$k$ Grothendieck problem which forces the ground state to be spread out instead of localized at a single point. This is a continuous analog of motivation previously given for the transverse-field Ising model \cite{Parekh_2023}, which can be interpreted as a version of classical \textsc{Max-Cut} with a term to encourage the ground state to be spread out.

There are additionally some examples coming from stochastic processes. Associated to any continuous-time Markov processes over a continuous state space is an infinitesimal generator, which is an essentially self-adjoint operator with respect to the inner product arising from the stationary measure. One can think of this operator as the Hamiltonian of a continuous-variable quantum system. While the ground energy is always 0 for such Hamiltonians, the ground state (here, related to the stationary measure) is still nontrivial and the moments output by the ncSoS SDP could potentially give some insight into it.

\subsection{The quantum rotor model}
The model is parameterized by an undirected graph $G = (V, E)$ on $n$ vertices, an integer $k \geq 1$, and two constants $a, b \in \mathbb{R}_{\geq 0}$. The Hilbert space is the tensor product $L^{2}(S^{k})^{\otimes n}$, and a pure state is a wavefunction $\psi : (S^{k - 1})^{n} \to \mathbb{C}$ mapping a length-$n$ sequence of $k$-dimensional unit vectors to an amplitude. The local Hamiltonian for the model is given by the formula
\[ H_{G,n,a,b} = a\sum _{i \in V} \Delta_{i} + b\sum _{(i, j) \in E} (2 + \vec{x}_{i} \cdot \vec{x}_{j}) \]
where $\Delta_{i}$ is the Laplace-Beltrami operator on the $i$-th factor, and $\vec{x}_{i}$ is a $k$-tuple of operators giving the position coordinates of the $i$-th unit vector. The Laplace-Beltrami operator on the sphere is an analog of the Laplacian on $\mathbb{R}^{k}$, and serves as a kinetic energy term. The constant 2 was chosen for technical reasons discussed after Theorem~\ref{thm:BOV}.

The quantum rotor model is related to several other models in quantum mechanics and quantum information, both exactly and approximately:
\begin{enumerate}
\item When $a = 0$, all terms in the Hamiltonian commute and it is a classical function. This is exactly the objective function of the classical rank-$k$ Grothendieck problem \cite{Briet_Oliveira_Vallentin_2014}. If additionally $k = 1$ the rank-$k$ Grothendieck problem is equivalent to classical \textsc{Max-Cut} (although the Laplace-Beltrami operator becomes 0).
\item When $k = 2$, the quantum rotor model can occur as a low-energy effective theory of superconducting Josephson junction arrays or bosons in optical lattices \cite{Vojta_Sknepnek_2006}.
\item When $k = 3$, the quantum rotor model can describe a bilayer quantum Heisenberg antiferromagnet or a double-layer quantum Hall ferromagnet \cite{Vojta_Sknepnek_2006, Sachdev_2000}
\item When $k = 3$, the quantum rotor model can be derived as an effective theory of a Heisenberg spin chain or 2-D Heisenberg model. This was used in Haldane's work on the existence of a mass gap (spectral gap in the thermodynamic limit) for integer and half-integer spin Heisenberg spin chains \cite{Haldane_1983a, Haldane_1983b, Haldane_1988, Shankar_Read_1990, Haldane_2016}.
\item When $k \geq 3$, the quantum rotor model has been studied as a toy model of non-abelian lattice gauge theory (in the Hamiltonian formalism) exhibiting many similar phase transitions and other phenomena \cite{Milsted_2016}.
\item The quantum rotor model is the sum of a 2-local potential and 1-local terms which encourage the ground state to be spread out. The 1-local terms have a purely classical interpretation in operations research in terms of a diversity constraint, which has previously been used to motivate the transverse-field Ising model \cite{Parekh_2023}.
\end{enumerate}

Previous work has done numerical experiments to study various algorithms and ans\"{a}tze for the quantum rotor model, including
\begin{itemize}
\item diffusion Monte Carlo \cite{Hetenyi_Berne_2001}
\item cluster Monte Carlo \cite{Alet_Sorensen_2003}
\item matrix product states \cite{Garcia_Cirac_Zoller_Kollath_Schollwock_Delft_2004, Iblisdir_Orus_Latorre_2007, Milsted_2016}
\item path integral ground state Monte Carlo \cite{Abolins_Zillich_Whaley_2011}
\item tensor-entanglement renormalization group \cite{LaPlaca_2014}
\item density matrix renormalization group \cite{Iouchtchenko_Roy_2018}
\item continuous-variable neural network quantum states \cite{Stokes_De_Veerapaneni_Carleo_2023}. 
\end{itemize}
Initial work on analyzing ncSoS for the rotor model was done by \cite{Hastings_2023}. To our knowledge, ours is the first result providing theoretical guarantees for finding the ground energy of the quantum rotor model.
The quantum rotor model has also been studied in a mean-field spin glass setting, with random Gaussian all-to-all interactions \cite{Ye_Sachdev_Read_1993}. Our results apply to arbitrarily weighted interaction graphs, including the quantum rotor spin glass in particular.

\subsection{Main result}
The main result of this paper is
\begin{theorem}[informal]
For $k$ sufficiently large, a level-1 ncSoS relaxation achieves an approximation ratio better than the best possible product state ratio for $H_{G,n,a,b}$.
\end{theorem}

A nontrivial amount of work is needed just to even write the quantum rotor model as a ncPOP,
which is in turn needed to write down the level-1 ncSoS relaxation -- Section~\ref{sec:quantization} is in fact devoted entirely to this.
Moreover, in this case there are multiple possible ways to write the quantum rotor model as a ncPOP, and different ones can be useful in different contexts. The ncSoS proof of a spherical uncertainty principle in Section~\ref{sec:spherical-uncertainty} uses a different parameterization than the one used in our SDP relaxation.
This issue is discussed in more detail in Section~\ref{sec:parameterizations}.

To bound the product state approximation ratio, we use the fact that expectations of tensor product observables factor over a product state, along with a spherical uncertainty principle \cite{Erb_2010} to constrain the possible expectation values of the kinetic and potential terms for a single site. We show that this and a few other uncertainty principles have low-degree ncSoS proofs, which may explain why the ncSoS relaxation ``knows about'' entangled states.

To bound the SDP approximation ratio, we define a ``rounding map'' from SDP solutions to true states. Using ideas from noncommutative geometry, we re-interpret the rounding maps for classical \textsc{Max-Cut} and the rank-$k$ Grothendieck problem as a classical channel, and use this to define an analogous quantum channel for the rotor model. This channel is applied to a bosonic Gaussian state whose covariance matrix is derived from the SDP solution, and the resulting state is likely to be efficiently preparable on a quantum computer. The analysis of the approximation ratio uses Wick's formula for the moments of a bosonic Gaussian and an expression for the channel in the Heisenberg picture.

\subsection{Open questions}
The main open question raised by this paper is whether this rounding strategy can be applied to other problems or to higher ncSoS levels. For the rotor model we use a relatively simple channel and the well-studied bosonic Gaussians, but other problems may require more complicated channels and states. Note that this strategy can still be used for finite-dimensional problems as long as the channel outputs a finite-dimensional state, and an interesting open question is if this can give better approximations in finite dimensions. For the rotor model in particular, it would be interesting to determine (1) if the SDP and parameterization in this paper is optimal, and (2) if our rounding scheme is optimal for this particular SDP.

\section{Technical overview}
\subsection{Problem}
We consider the \emph{quantum rotor model}. The model is parameterized by an undirected graph $G = (V, E)$ on $n$ vertices, an integer $k \geq 1$, and two constants $a, b \in \mathbb{R}_{\geq 0}$. The Hilbert space is the tensor product $L^{2}(S^{k - 1})^{\otimes n}$, and a pure state is a wavefunction $\psi : (S^{k - 1})^{n} \to \mathbb{C}$ mapping a length-$n$ sequence of $k$-dimensional unit vectors to an amplitude. The local Hamiltonian for the model is given by the formula
\[ H_{G,n,a,b} = a\sum _{i \in V} \Delta_{i} + b\sum _{(i, j) \in E(G)} (1 + x_{i} \cdot x_{k}) \]
where $\Delta_{i}$ is the Laplace-Beltrami operator on the $i$-th tensor factor, and $x_{i}$ is a $k$-tuple of operators giving the position coordinates of the $i$-th unit vector. The model has a global on-site $O(k)$ symmetry, since the orthogonal group preserves inner products and the Laplace-Betrami operator on $S^{k - 1}$ with the standard metric is rotationally invariant. The main problem we consider is calculating the ground-energy of the Hamiltonian. (Essential self-adjointness of this Hamiltonian follows from essential self-adjointness of the Laplace-Beltrami operator on the sphere and the Kato-Rellich theorem.)

When $a = 0$ and $b > 0$ the problem reduces to a classical Hamiltonian and was considered in \cite{Briet_Oliveira_Vallentin_2014}, who called it the ``rank-$k$ Grothendieck problem''. This problem is not known to be NP-hard for constant $k$, but in \cite{Hwang_Neeman_Parekh_Thompson_Wright_2023} it was conjectured that it is actually NP-hard. When $a > 0$ and $b = 0$ the ground state wavefunction is constant and the ground energy is 0. Quantum behavior arises when both $a, b > 0$, and we consider the algorithmic problem of calculating the ground energy in this regime. General facts strongly suggest that this problem is in $\mathsf{StoqMA}$, but to our knowledge it is not known to be $\mathsf{StoqMA}$-hard.

\subsection{Approach}
The approach is based on a hierarchy of semidefinite programming relaxations for noncommutative polynomial optimization problems. At a non-rigorous level, the Hamiltonian for the rotor model is a special case of a nc polynomial optimization problem.

\subsubsection{Approach to writing down a ncSoS relaxation}\label{sec:parameterizations}
As noted in the introduction, this work uses multiple different ``presentations'' of the algebra $B(L^{2}(S^{k - 1}))$ of operators on the Hilbert space $L^{2}(S^{k - 1})$, including one for the ncSoS relaxation and another for the ncSoS proof of the spherical uncertainty principle. The ambiguity arises in defining the appropriate analog of momentum operators when the configuration space is a sphere. When the configuration space is $\mathbb{R}^{n}$, one can define the momentum operators in a canonical way by taking the $i$-th momentum operator $p_{i}$ to be the directional derivative in the direction $e_{i}$ (up to an imaginary scaling factor). This works because the smooth manifold $\mathbb{R}^{n}$ is \textit{parallelizable} -- that is, the tangent bundle $T\mathbb{R}^{n} \cong \mathbb{R}^{n} \oplus \mathbb{R}^{n}$ has $n$ linearly independent smooth sections. (A section of the tangent bundle of a manifold $M$ is the same as a vector field over $M$.) More generally, any Lie group is parallelizable (since one can take a basis of the Lie algebra and define smooth vector fields by translation), and in this case an analogous set of momentum operators can be defined by taking a directional derivative operator along each vector field.

It was a long-standing open problem to determine if the sphere $S^{k - 1}$ is parallelizable, which was eventually solved by Adams in 1962 \cite{Adams_1962} using a number of involved tools from algebraic topology, including topological $K$-theory and generalized cohomology, as well as stable homotopy theory. Unfortunately, it turns out that $S^{k - 1}$ is only parallelizable when $k \in \{1, 2, 4, 8\}$. Thus one is forced to consider an overcomplete set of linearly dependent vector fields on $S^{k - 1}$, which in turn define an overcomplete set of momentum operators.

A natural approach to finding a spanning set of vector fields on a sphere is to note that $S^{k - 1}$ is immersed in $\mathbb{R}^{k}$, which is parallelizable. Thus there is a set of $k$ (overcomplete) vector fields given by restricting the standard $k$ vector fields on $\mathbb{R}^{n}$ to $S^{k - 1}$ and projecting onto $TS^{k - 1}$ (using the standard Riemannian metric on $\mathbb{R}^{k}$). These are the vector fields used to define our parameterization in Section~\ref{sec:quantization}, and the SDP and rounding algorithm.

In order to determine the full set of commutation relations for the corresponding momentum operators we need to calculate their adjoints, which involves a term coming from the divergence of the vector field taken with respect to the standard (round) Riemannian metric on $S^{k - 1}$ (given by restricting the standard metric from $\mathbb{R}^{k}$).
The Riemannian divergence is usually defined only in local coordinates and would normally require a lengthy calculation with respect to an explicit atlas of $S^{k - 1}$. For spheres we are able to prove Lemma~\ref{lem:divergence} which relates the Riemannian divergence of a vector field over $S^{k - 1}$ to the usual divergence of a 0-homogeneous extension of that vector field to $\mathbb{R}^{k - 1} \setminus \{0\}$, and thus reduce the divergence calculation to standard multivariable calculus over $\mathbb{R}^{k}$. This is also used to write the Laplacian $\Delta = \operatorname{div} \circ \nabla$ over $S^{k - 1}$ in terms of these momentum operators.

For our ncSoS proof of the spherical uncertainty principle in Section~\ref{sec:spherical-uncertainty}, we use a different set of momentum operators. These arise from differentiating the action of the special orthogonal group $SO(k)$ on $S^{k - 1}$ to get an action of the Lie algebra $\mathfrak{so}(k)$ on $L^{2}(S^{k - 1})$. This Lie algebra action defines a set of first-order differential operators and thus a set of vector fields. This set of vector fields has cardinality $\binom{k}{2}$ as opposed to $k$ for the set used in our SDP and rounding algorithm, and thus would give a larger and more inefficient SDP. However, these vector fields seem to have slightly better symmetry properties which are useful for proving the spherical uncertainty principle. It would be interesting to see if there is a proof at low degrees using the presentation from the SDP.

\subsubsection{Approach to rounding the SDP solution}
To prove that the value of the SDP solution approximates the value of the true solution, the typical strategy is to give two maps: one which takes a true solution and gives an SDP solution, and one the other way which takes an SDP solution and gives a true solution. The first map is typically so clear from context that it is not explicitly considered in prior literature, but in principle it could be highly non-trivial. In prior literature (and here) it also typically preserves the value exactly and so the SDP is called a relaxation, but in principle there could be interesting examples where the it does not preserve the value exactly.

The second map is usually more difficult to understand. In prior literature it is called the ``rounding map'' by analogy with linear programming relaxations of integer programming problems, where it is given by rounding a real vector to the nearest integer vector. When the map is efficiently computable (possibly using randomness), it is also called a ``rounding algorithm.'' The most common choice of parameter for measuring the quality of a rounding map usually is a bound on the ratio between the value of the given SDP solution and the value of the rounded solution. This bound is often proven term-by-term when the objective function is a sum of local terms. The main result of \cite{Briet_Oliveira_Vallentin_2014} is the analysis of a rounding map for the rank-$k$ Grothendieck problem.

Typically in the classical case, the rounding algorithm is randomized and involves sampling from a distribution over a larger domain whose parameters are determined by the SDP solution, followed by step which ``rounds'' the solution back onto the domain of the original problem. For classical \textsc{Max-Cut} and the rank-$k$ Grothendieck problem, the distribution is a Gaussian whose covariance is the SDP moment matrix and the rounding map takes the angular part of a point in $\mathbb{R}^{k}$ (or sign when $k = 1$). One can interpret the second step as a deterministic classical channel acting on a Gaussian distribution (with the precise correspondence given by a duality between spaces and algebras of functions). For the quantum rotor model, we define a quantum channel which is applied to a bosonic Gaussian state.

In order to derive either the Goemans-Williamson or BOV SDP from the classical SoS framework, one needs to use a symmetry reduction argument on the SDP coming from SoS. For both algorithms this symmetry reduction is not particularly important, but it becomes critical in our rounding for the quantum rotor model. The reason is that a bosonic Gaussian states requires very particular constraints on it covariance matrix (namely that it respects the canonical commutation relations), but the operators we define on $L^{2}(S^{k - 1})$ do not satisfy the exact same commutation relations and thus the degree-2 SDP moment matrix cannot be used directly as the covariance of the bosonic Gaussian. However, after doing a careful symmetry reduction step the form of the moment matrix does become very similar to that of a bosonic Gaussian covariance matrix, and then a simple rescaling allows the moment matrix to be used to define a bosonic Gaussian. This rescaling factor ends up being used to bound the final approximation ratio. These calculations are described in Section~\ref{sec:symmetry}.

For similar reasons as in the approximation algorithm, the SoS proofs we give of uncertainty principles also need a symmetry reduction step. Doing the symmetry reduction typically requires some basic tools from representation theory, including simple arguments involving Schur's lemma and calculations of small plethysms and tensor product multiplicities (similar to Clebsch-Gordan coefficients in the physics literature).

Our quantum channel is easiest to define in the Schr\"{o}dinger picture, by tracing out the ``radial subsystem'' of the bosonic Gaussian, but the analysis uses the Heisenberg picture (related to the algebra-geometry duality mentioned earlier). Calculating the ``amplitudes'' of the density operator from the channel acting on the bosonic Gaussian would involve a complicated integral, but using the Heisenberg picture one can have the channel instead act on observables. The expressions for our channel acting on the terms in the quantum rotor Hamiltonian turn out to be relatively simple, and their expectation values can be calculated with respect to the Wigner function of the bosonic Gaussian (which is just a classical Gaussian distribution). The overall calculation is done in Section~\ref{sec:rounding-analysis}.

\subsection{Comparison to existing approaches for rounding degree-2 ncSoS}
The classical Gaussian-based approaches to rounding degree-2 SoS relaxations \cite{Goemans_Williamson_1995, Briet_Oliveira_Vallentin_2014} perform the following steps (illustrated here for the rank-$k$ Grothendieck problem):
\begin{enumerate}
\item Sample a matrix $P \in \mathbb{R}^{k \times n}$ whose entries are independent standard Gaussian random variables.
\item Apply $P$ to the vectors $v_{1}, \dots, v_{n}$ obtained from the SDP, which are the columns of the Cholesky factor of the moment matrix.
\item Round $Pv_{1}, \dots, Pv_{n}$ to $S^{k - 1}$ using the function $f(x) = x/|x|$.
\end{enumerate}
Traditionally, steps 2 and 3 are combined. In this way the rounding algorithm is described as sampling a random hyperplane (or random subspace when $k \geq 2$) and determining which side of the hyperplane each SDP vector falls on (or the direction each SDP vector projects to inside the random subspace). Alternatively, the same rounding algorithm can be equivalently described with steps 1 and 2 grouped together, which amounts to sampling a multivariate Gaussian of dimension $k \times n$ whose covariance matrix is given by the Kronecker product $M \otimes I_{k}$ where $M$ is the SDP moment matrix.

In our approach to a quantum analog of Gaussian rounding, only the second type of description works. The reason is that the equivalence between the two descriptions requires being able to transform a multivariate isotropic Gaussian into an arbitrary Gaussian by applying a linear transformation. For bosonic Gaussians, it is only valid to use Bogoliubov transformations (analogous to linear canonical transformations in classical mechanics) since they preserve the canonical commutation relations (CCR). Different bosonic Gaussians have different ``distributions of classical mixture,'' which is preserved by Bogoliubov transformations, and so not all bosonic Gaussians are equivalent in this sense. Our approach will alway produces a valid covariance matrix satisfying the CCR, but it is not necessarily clear what kind of classical mixture it contains.

Hypothetically, if the first type of description worked then one could very loosely interpret the initial step as ``preparing a family of subspaces in superposition'' (instead of sampling a random subspace), although this analogy will immediately break down once one tries to make any part of it precise. Initial approaches to \textsc{Quantum Max-Cut} rounded using a classically random subspace or projection \cite{Gharibian_Parekh_2019}, limiting the amount of entanglement that could be in the final state. Later approaches then built on this approach by adding entanglement to this type of construction, sometimes by feeding the low-entanglement state into an entangling quantum circuit. The bosonic Gaussians we use are entangled from the beginning, leading to some nontrivial amount of entanglement in the final rounded state.

\section{Warm-up: BOV SDP and rounding via SoS}
\subsection{The rank-\texorpdfstring{$k$}{k} Grothendieck problem}
When $a = 0$ in the quantum rotor model, the Hamiltonian is equivalent to the classical \textbf{rank-$k$ Grothendieck problem} whose input is a graph $G = (V, E)$ on $n$ vertices and whose output is
\[ \min _{x \in (S^{k - 1})^{n}} \sum _{(i, j) \in E} (2 + \vec{x}_{i} \cdot \vec{x}_{j}). \]
The reason for the name is that the problem is equivalent to
\begin{align*}
\min \sum _{(i, j) \in E} 2 + A_{ij} && \text{s.t.} \begin{aligned}[t]
A &\succeq 0 \\
A_{ii} &= 1 \\
\operatorname{rank}(A) &= k.
\end{aligned}
\end{align*}
When $k = 1$, the problem is equivalent to classical \textsc{Max-Cut}.

\subsection{BOV SDP and symmetry reduction}
In \cite{Briet_Oliveira_Vallentin_2014} the SDP is presented directly, but it is actually a special case of the classical SoS hierarchy at level 1 after doing a symmetry reduction on the SDP. To describe the unsymmetrized SDP, let
\begin{align*}
R &= \mathbb{C}[x_{v,i} : v \in V(G), i \in [k]] \\
W &= \operatorname{span} (\{1\} \cup \{x_{v,i} : v \in V(G), i \in [k]\}).
\end{align*}
For every $v \in V(G)$ define an action of the orthogonal group $O(k)$ on $\operatorname{span} \{x_{v, i} : i \in [k]\}$ by taking it to be a copy of the standard representation. The action extends to $R$ and $W$ in the natural way.

The SoS moment matrix is a symmetric bilinear form $M : W \times W \to \mathbb{R}$. Denote by $M_{v, w} \in \mathbb{C}^{k \times k}$ the $k \times k$ block of $M$ corresponding to vertices $v, w$ and $M_{v, 1} = M_{1, v}^{T}$ the $k \times 1$ block corresponding to $v$ and the constant monomial $1$. In terms of the pseudoexpectation, we have
\begin{align*}
M_{v,w} &= \tilde{E}[x_{v} x_{w}^{T}] \\
M_{v, 1} &= \tilde{E}[x_{v}].
\end{align*}
The SoS relaxation at level 1 is
\begin{align*}
&& \max \sum _{(v, w) \in E(G)} (1 - \operatorname{tr} M_{v,w}) \\
\text{s.t.} && M &\succeq 0 \\
&& \operatorname{tr} M_{v,v} &= 1.
\end{align*}

This problem is convex and invariant under the $O(k)$ action. Thus without loss of generality, any feasible solution $M$ can be assumed to be invariant under the action (following the symmetry reduction for SoS described in \cite{Gatermann_Parrilo_2004}). By Schur's lemma there is a matrix $M' \in \mathbb{R}^{V(G) \times V(G)}$ such that $M_{v, w} = M'_{v,w}(\frac{1}{k}I_{k})$ and $M_{v, 1} = 0$ for all $v, w \in V(G)$. Thus $M = 1 \oplus M' \otimes \frac{1}{k}I_{k}$.

The BOV SDP optimizes directly over $M'$. This is useful partly because the size of $M$ depends on $k$, but $M'$ is always an $n \times n$ matrix and its size does not depend on $k$.

\subsection{BOV rounding}
To describe the rounding map from \cite{Briet_Oliveira_Vallentin_2014}, it is more convenient to redefine $M = M' \otimes I_{k}$. Let $f : \mathbb{R}^{k} \to S^{k - 1}$ be given by
\[ f(x) = \begin{cases} x/|x| & x \neq 0 \\
\text{arbitrary} & x = 0. \end{cases} \]

The rounding takes a sample $(y_{v})_{v \in V(G)} \sim N(0, M)$ and outputs the sequence of vectors $(f(y_{v}))_{v \in V(G)}$. In expectation, the value of the rounded solution is
\begin{align*}
E_{y}\left[ \sum _{(v, w) \in E(G)} (1 - f(y_{v}) \cdot f(y_{w})) \right] &= \sum _{(v, w) \in E(G)} E_{(y_{v}, y_{w})}[1 - f(y_{v}) \cdot f(y_{w})].
\end{align*}
In the right-hand side we have used the fact that the marginal of a Gaussian is a Gaussian. The analysis of the approximation ratio uses the fact that each term depends only on $M'_{v,w}$. The standard approach to analyzing randomized rounding uses this fact along with linearity of expectation to reduce the calculation of the approximation ratio to a single finite-dimensional optimization problem, which can then be solved numerically. We state these ideas more precisely in the proof of the following theorem.

\begin{theorem}[\cite{Briet_Oliveira_Vallentin_2014}]\label{thm:BOV}
In expectation, the BOV rounding achieves a constant approximation ratio $\alpha_{BOV,k}$.
\end{theorem}
\begin{proof}
For notational convenience, define
\[ A \coloneqq \begin{bmatrix}1 & M'_{v,w} \\ M'_{v, w} & 1\end{bmatrix}. \]
Using linearity of expectation, the expected value of the output solution is
\begin{align*}
E_{y \sim N(0, M)}\left[ \sum _{(v, w) \in E(G)} (2 + f(y_{v}) \cdot f(y_{w})) \right] &= \sum _{(v, w) \in E(G)} E_{(y_{v}, y_{w})}[2 + f(y_{v}) \cdot f(y_{w})] \\
\shortintertext{and the fact that the marginal of a Gaussian is Gaussian gives}
&= \sum _{(v, w) \in E(G)} E_{(y_{v}, y_{w}) \sim N(0, A \otimes \frac{1}{k}I_{k})}[2 + f(y_{v}) \cdot f(y_{w})] \\
&\leq \left( \max _{t \in [-1, 1]} g_{BOV,k}(t) \right) \sum _{(v, w) \in E(G)} (2 + M'_{v,w})
\end{align*}
where we define
\[ g_{BOV,k}(M'_{v,w}) \coloneqq \left(E_{(y_{v}, y_{w}) \sim N(0, A \otimes \frac{1}{k}I_{k})}[2 + f(y_{v}) \cdot f(y_{w})]\right)\frac{1}{2 + M'_{v,w}}. \]
Thus the approximation ratio is
\[ \alpha_{BOV,k} \coloneqq \max _{t \in [-1, 1]} g_{BOV,k}(t). \qedhere \]
\end{proof}

Note that Theorem~\ref{thm:BOV} is stated differently from in \cite{Briet_Oliveira_Vallentin_2014}, since we have phrased the rank-$k$ Grothendieck problem as a minimization problem for similarity to the quantum rotor model as opposed to a maximization problem. In order to achieve a constant approximation ratio the function $g_{BOV,k}$ must be bounded above on $[-1, 1]$, but if the constant in the objective was chosen to be 1 this would not be true. We have chosen $2$ for simplicity, which does change the approximation ratio, but changes it in the same way for both the product state and SDP approximation ratios. For any reasonable definition it will still hold that degree-2 ncSoS outperforms the best product state approximation.

Since our result holds for sufficiently large $k$, we will need to analyze the BOV algorithm in this regime.

\begin{lemma}\label{lem:gauss-k}
Suppose $x \sim N(0, \frac{1}{k}I_{k})$. Then
\[ \lim _{k \to \infty} E[|x/|x| - x|^{2}] = 0. \]
\end{lemma}
\begin{proof}
We have
\begin{align*}
E[|x/|x| - x|^{2}] &= E[(x/|x|) \cdot (x/|x|) - 2x\cdot(x/|x|) + x \cdot x] \\
&= E[1] - E[2|x|^{2}/|x|] + E[x \cdot x] \\
&= 2(1 - E[|x|]) \\
&= 2\left(1 - \frac{\Gamma(\frac{k + 1}{2})}{\Gamma(\frac{k}{2})}\sqrt{\frac{2}{k}} \right)
\end{align*}
using a standard formula for the mean of a chi distribution. When $k = 2l$ for $l \in \mathbb{N}$, we have
\begin{align*}
\Gamma\left(\frac{k}{2}\right) &= (l - 1)! \\
\Gamma\left(\frac{k + 1}{2}\right) &= \frac{(2l)!}{4^{l}l!}\sqrt{\pi}.
\end{align*}
Thus
\begin{align*}
\frac{\Gamma(\frac{k + 1}{2})}{\Gamma(\frac{k}{2}} &= \sqrt{\pi} 4^{-l}l\binom{2l}{l} \\
&= \sqrt{\pi} 4^{-l} l\frac{4^{l}}{\sqrt{\pi l}}(1 - O(1/l)) \\
&= \sqrt{k/2}(1 - O(1/k)).
\end{align*}
When $k = 2l + 1$ is odd, we have
\begin{align*}
\frac{\Gamma(\frac{k + 1}{2})}{\Gamma(\frac{k}{2})} &= l!\left( \frac{(2l)!}{4^{l}l!}\sqrt{\pi} \right)^{-1} \\
&= \left(\frac{\sqrt{\pi}}{4^{l}} \binom{2l}{l} \right)^{-1} \\
&= \left( \frac{\sqrt{\pi}}{4^{l}} \frac{4^{l}}{\sqrt{\pi l}}(1 - O(1/l)) \right)^{-1} \\
&= \sqrt{k/2}(1 - O(1/k)).
\end{align*}
\end{proof}

\begin{lemma}\label{lem:BOV-k}
Suppose $(x, y) \sim N(0, A \otimes \frac{1}{k}I_{k})$ where $A = \begin{bmatrix}1 & a \\ a & 1 \end{bmatrix}$. Then
\[ \lim _{k \to \infty} E[(x/|x|) \cdot (y/|y|)] = a. \]
\end{lemma}
\begin{proof}
We have
\begin{align*}
E[(x/|x|) \cdot (y/|y|)] &= E[(x/|x| - x + x) \cdot (y/|y| - y + y)] \\
&= E[x \cdot y + (x/|x| - x) \cdot (y/|y| - y) + (x/|x| - x)\cdot y + x \cdot (y/|y| - y)] \\
&= a + E[(x/|x| - x) \cdot (y/|y| - y)] + E[(x/|x| - x)\cdot y] + E[x \cdot (y/|y| - y)].
\end{align*}
Each of the last three terms converges to 0, since using Cauchy-Schwarz on the space $L^{2}(\mathbb{R}^{2k}, \mathbb{R}^{k})$ of vector-valued functions gives
\begin{align*}
|E[(x/|x| - x) \cdot (y/|y| - y)]|^{2} &\leq E[|x/|x| - x|^{2}]E[|y/|y| - y|^{2}] \\
|E[(x/|x| - x)\cdot y]|^{2} &\leq E[|x/|x| - x|^{2}]E[|y|^{2}] = E[|x/|x| - x|^{2}] \\
|E[x \cdot (y/|y| - y)]|^{2} &\leq E[|x|^{2}]E[|y/|y| - y|^{2}] = E[|y/|y| - y|^{2}]
\end{align*}
and the right-hand sides all converge to 0 by Lemma~\ref{lem:gauss-k}.
\end{proof}

\begin{proposition}\label{prop:BOVconv}
We have $\lim _{k \to \infty} \alpha_{BOV,k} = 1$.
\end{proposition}
\begin{proof}
The functions $g_{BOV,k}$ are all bounded and continuous on $[-1, 1]$. By Lemma~\ref{lem:BOV-k} we know that $g_{BOV,k}$ converges pointwise to the constant function 1 on $[-1, 1]$, which is also continuous. Thus the convergence is in fact uniform, which implies that
\begin{align*}
\lim _{k \to \infty} \alpha_{BOV,k} &= \lim _{k \to \infty} \max _{t \in [-1, 1]} g_{BOV,k}(t) \\
&= \lim _{k \to \infty} 1 \\
&= 1. \qedhere
\end{align*}
\end{proof}

\section{Phase space and quantization on the sphere}\label{sec:quantization}
In order to describe the noncommutative SoS relaxation of the quantum rotor model, we must first express the Hamiltonian in terms of more basic operators satisfying certain relations. In the case of the rank-$k$ Grothendieck problem where the domain was a product of spheres, these were the variables $\{x_{v,i}\}_{v \in V(G), i \in [k]}$ satisfying the relations that all variables commute with each other and $\sum _{i = 1} ^{k} x_{v,i}^{2} = 1$ for all $v \in V(G)$. 

For the quantum rotor model, the set of variables will include $\{x_{v,i}\}$ but will also include additional variables meant to represent first-order differential operators (that is, vector fields) on the sphere, which in physics correspond to momentum coordinates. These variables will not all commute with each other and so the set of relations will also include commutation relations between certain variables.

It will be instructive to first describe the classical phase space whose configuration space is a sphere, since its Poisson brackets will essentially determine the commutation relations of the quantum operators.

\subsection{Phase space}
The classical phase space for an unconstrained particle moving in $\mathbb{R}^{k}$ is $T^{*}\mathbb{R}^{k} \cong \mathbb{R}^{k} \oplus \mathbb{R}^{k}$. The typical notation for a point in phase space is $(x, p)$ or $(x_{1}, \dots, x_{k}, p_{1}, \dots, p_{k})$ where $x$ is the position and $p$ is the momentum. Any cotangent bundle comes with a canonical symplectic form, which can be used to treat it as a phase space and define classical Hamiltonian mechanics on the cotangent bundle. For $T^{*}\mathbb{R}^{k}$, this symplectic form is the constant 2-form $\sum _{i = 1} ^{n} dx_{i} \wedge dp_{i}$.

For a particle constrained to the sphere $S^{k - 1}$, the phase space is $T^{*}S^{k - 1}$. Viewing $S^{k - 1}$ as being isometrically embedded into $\mathbb{R}^{n}$ as Riemannian manifolds, we have $T_{x}S^{k - 1} \cong \{x\}^{\perp}$. Thus
\[T^{*}S^{k - 1} \cong \{(x, p) : |x|^{2} = 1, x \cdot p = 0\} \subseteq T^{*}\mathbb{R}^{n}. \]
The symplectic form on $T^{*}S^{k - 1}$ comes from restricting the one on $T^{*}\mathbb{R}^{k}$.

For convenience, define new vector fields
\begin{align*}
\tilde{p}_{i} &\coloneqq \nabla x_{i} \\
&= \sum _{j = 1} ^{k} (|x|^{2}\delta_{ij} - x_{i}x_{j})e_{j}.
\end{align*}
Direct calculation shows that $\tilde{p}_{i}(x) \in \{x\}^{\perp}$ for $x \in S^{k - 1}$, so these define vector fields on $S^{k - 1}$. These also define functions on $T^{*}S^{k - 1}$ by
\[ p'_{i}(x, p) \coloneqq \langle p, \tilde{p}_{i}(x) \rangle \]
using the Riemannian structure on $S^{k - 1}$. The corresponding Poisson brackets on $T^{*}S^{k - 1}$ (also called Dirac brackets) are
\begin{align*}
\{x_{i}, x_{j}\} &= 0 \\
\{\tilde{p}_{j}, x_{i}\} &= \delta_{ij} - x_{i}x_{j} \\
\{\tilde{p}_{i}, \tilde{p}_{j}\} &= x_{j}\tilde{p}_{i} - x_{i}\tilde{p}_{j}.
\end{align*}

\subsection{Quantization}
To quantize the classical observables, take the Hilbert space $H = L^{2}(S^{k - 1})$. To quantize position, define $\hat{x}_{i}$ to be the multiplication operator of the function $S^{k - 1} \ni x \mapsto x_{i}$, which is a bounded operator (bounded by 1) on $H$.

To quantize momentum, define
\[ (\hat{p}_{i}f)(x) \coloneqq \partial_{\tilde{p}_{i}(x)}f(x) \]
for $f \in C^{\infty}(S^{k - 1})$, where the derivative is a directional derivative. This operator is unbounded and densely defined, and its domain contains $C^{\infty}(S^{k - 1})$. By the divergence theorem, the adjoint is
\[ \hat{p}_{i}^{*} = -\hat{p}_{i} + \operatorname{div}(\tilde{p}_{i}). \]
To calculate the divergence, we first prove the following lemma.

\begin{lemma}\label{lem:divergence}
Let $M$ be an oriented Riemannian manifold, $S \subseteq M$ be an oriented Riemannian submanifold, $p : N \to S$ be the normal bundle of $S$ in $M$, and $j : N \to M$ be an embedding such that $j(N)$ is a tubular neighborhood of $S$. Suppose $s : S \to TS$ is a vector field and let $s' : j(N) \to TM$ be given by $s' = s \circ p \circ j^{-1}$. Then $\operatorname{div} s'(x) = \operatorname{div} s(x)$ for all $x \in S$.
\end{lemma}
\begin{proof}
Let $x \in S$ and $y_{1}, \dots, y_{n}$ be local coordinates on an open neighborhood $U \ni x$ in $M$ such that $y_{k + 1}, \dots, y_{n}$ vanish on $S \cap U$. It suffices to prove the lemma in any such chart. Let $(v_{1}, \dots, v_{n})$ be a positively oriented local frame such that $(v_{1}, \dots, v_{k})$ is a positively oriented local frame of $S$ and are all orthogonal to $v_{k + 1}, \dots, v_{n}$ when restricted to $S \cap U$. In the given chart, we have $g_{M} = g_{S} \oplus g'$ and $\det g_{M} = (\det g_{S})(\det g')$. Then the Riemannian volume forms satisfy
\[ \omega_{M} = \sqrt{|g_{M}|} (dv_{1}) \wedge \cdots \wedge (dv_{n}) = \omega_{S} \wedge \omega' \]
for some $\omega'$. We know $\omega'(s') = 0$ by orthogonality, so
\[ (\operatorname{div} s')\omega_{M} = d(i_{s'}\omega_{M}) = d(i_{s}\omega_{S} \wedge \omega') = (\operatorname{div} s)\omega_{S} \wedge \omega' + (-1)^{k}i_{s}\omega_{S} \wedge d\omega'. \]
We have
\[ d\omega' = d(\sqrt{|g'|}(dv_{k + 1}) \wedge \dots \wedge (dv_{n})) = 0 \]
so $\operatorname{div} s' = \operatorname{div} s$ in the given chart.
\end{proof}

The particular case used for the rotor model is where $M = \mathbb{R}^{k}$ and the submanifold is $S^{k - 1}$. In that case, the tubular neighborhood of $S^{k - 1}$ is $\mathbb{R}^{k} \setminus \{0\}$ and $(p \circ j^{-1})(x) = x/|x|$. Then
\[ \hat{p}_{i}(x/|x|) = \left(\delta_{ij} - \frac{x_{i}x_{j}}{|x|^{2}}\right)_{j \in [k + 1]} \]
and
\begin{align*}
\operatorname{div} \hat{p}_{i}(x) &= \sum _{j = 1} ^{k + 1} \partial_{j} \frac{-x_{i}x_{j}}{|x|^{2}} \\
&= \frac{-2x_{i}(|x|^{2} - x_{i}^{2})}{|x|^{4}} - \sum _{j \in [k + 1] \setminus \{i\}} \frac{x_{i}(|x|^{2} - 2x_{j}^{2})}{|x|^{4}} \\
&= \frac{2x_{i}}{|x|^{4}}\sum _{j = 1} ^{k + 1} x_{j}^{2} - (k + 1)x_{i}\frac{|x|^{2}}{|x|^{4}} \\
&= -(k - 1)\frac{x_{i}}{|x|^{2}}.
\end{align*}
In particular, $\operatorname{div} \hat{p}_{i}(x) = -(k - 1)x_{i}$ for $x \in S^{k - 1}$.

\begin{lemma}
Let $f : S^{k - 1} \to \mathbb{R}$ and $g : \mathbb{R}^{k} \to \mathbb{R}$ be given by $g(x) = f(x/|x|)$. Then $(\partial_{i} g)\upharpoonright_{S^{k - 1}} = \hat{p}_{i}f$.
\end{lemma}
\begin{proof}
Let $i \in [k + 1]$. Recall the identification $T_{x}S^{k} = \{x\}^{\perp}$. Then
\[ \partial_{i} \frac{x_{j}}{|x|} = |x|^{-3} (|x|^{2}\delta_{ij} - x_{i}x_{j}) \]
so
\begin{align*}
(\partial_{i} g) \upharpoonright_{S^{k}} &= (df(|x|^{-3} (|x|^{2}\delta_{ij} - x_{i}x_{j}))_{j \in [k+1]})\upharpoonright_{S^{k}} \\
&= df((|x|^{2}\delta_{ij} - x_{i}x_{j})_{j \in [k+1]}) \\
&= df(\tilde{p}_{i}) \\
&= \hat{p}_{i}f. \qedhere
\end{align*}
\end{proof}

The classical Dirac brackets turn to out to give some of the correct commutators for the quantized operators without needing any higher-order corrections.

\begin{proposition}
The operators $\{\hat{x}_{i}\}$ and $\{\hat{p}_{i}\}$ satisfy the relations
\begin{align*}
[\hat{x}_{i}, \hat{x}_{j}] &= 0 \\
[\hat{p}_{j}, \hat{x}_{i}] &= \delta_{ij} - \hat{x}_{i}\hat{x}_{j} \\
[\hat{p}_{i}, \hat{p}_{j}] &= \hat{x}_{j}\hat{p}_{i} - \hat{x}_{i}\hat{p}_{j}
\shortintertext{and}
\sum _{i = 1} ^{k} \hat{x}_{i}\hat{p}_{i} &= 0.
\end{align*}
\end{proposition}
\begin{proof}
The commutation of $\hat{x}_{i}$ and $\hat{x}_{j}$ is clear. For the second, we have
\begin{align*}
\hat{p}_{i}(\hat{x}_{j}f) &= \partial_{i} (x \mapsto f(x/|x|)x_{j}/|x|)\upharpoonright_{S^{k-1}} \\
&= (\partial_{i} (x \mapsto f(x/|x|)))\upharpoonright_{S^{k-1}} x_{j} + f \cdot (\partial_{i} (x_{j}/|x|))\upharpoonright_{S^{k-1}} \\
&= (\hat{p}_{i}f)\hat{x}_{j} + f\cdot \left( \frac{\delta_{ij}|x|^{2} - x_{i}x_{j}}{|x|^{3}} \right)\upharpoonright_{S^{k-1}} \\
&= (\hat{p}_{i}f)\hat{x}_{j} + (\delta_{ij} - \hat{x}_{i}\hat{x}_{j})f.
\end{align*}
For the third, assume $i \neq j$. Let $g(x) = f(x/|x|)$. Then
\begin{align*}
\hat{p}_{i} f &= \left(\sum _{l = 1} ^{k} (\partial_{l} g)(x/|x|) \cdot (\delta_{il} - |x|^{-2}x_{i}x_{l})\right)\upharpoonright_{S^{k - 1}}
\shortintertext{and}
\partial_{j} ((\hat{p}_{i} f)(x/|x|)) &= \sum _{l = 1} ^{k} \partial_{j}((\partial_{l} g)(x/|x|) \cdot (\delta_{il} - |x|^{-2}x_{i}x_{l})) \\
&= \sum _{l = 1} ^{k} \partial_{j}((\partial_{l} g)(x/|x|)) \cdot (\delta_{il} - |x|^{-2}x_{i}x_{l})) \\
&\phantom{{}={}} + \sum _{l = 1} ^{k} (\partial_{l} g)(x/|x|) \cdot \partial_{j} (\delta_{il} - |x|^{-2}x_{i}x_{l})).
\end{align*}
Thus
\begin{align*}
&\phantom{{}={}} \partial_{j}((\hat{p}_{i}f)(x/|x|)) - \partial_{i}((\hat{p}_{j}f)(x/|x|)) \\
&= \sum _{l = 1} ^{k} (\partial_{l} g)(x/|x|) \cdot (\partial_{j}(\delta_{il} - |x|^{-2}x_{i}x_{l}) - \partial_{i}(\delta_{jl} - |x|^{-2}x_{j}x_{l})) \\
&= \sum _{l \in [k] \setminus \{i, j\}} (\partial_{l} g)(x/|x|) \cdot (2x_{i}x_{j}x_{l}|x|^{-4} - 2x_{i}x_{j}x_{l}|x|^{-4}) \\
&\phantom{{}={}} + (\partial_{j} g)(x/|x|) \cdot (x_{i}(|x|^{2} - 2x_{j}^{2})|x|^{-4} - 2x_{i}x_{j}^{2}|x|^{-4}) \\
&\phantom{{}={}} + (\partial_{i} g)(x/|x|) \cdot (2x_{i}^{2}x_{j}|x|^{-4} - x_{j}(|x|^{2} - 2x_{i}^{2})|x|^{-4}) \\
&=(\partial_{j} g)(x/|x|) \cdot (x_{i}|x|^{-4}) - (\partial_{i} g)(x/|x|) \cdot (x_{j} |x|^{-4})
\end{align*}
and restricting to $S^{k - 1}$ gives the desired identity. For the last, we have
\begin{align*}
0 &= \nabla(1) \\
&= \nabla\left(\sum _{i = 1} ^{k + 1} x_{i}^{2}\right) \\
&= 2\sum _{i = 1} ^{k + 1} x_{i}\nabla (x_{i}) \\
&= 2 \sum _{i = 1} ^{k + 1} x_{i} \tilde{p}_{i}. \qedhere
\end{align*}
\end{proof}

Since the operators $\hat{p}_{i}$ are not symmetric, we will define symmetric operators by
\[ \hat{q}_{i} \coloneqq i\hat{p}_{i} + \frac{i}{2} \operatorname{div} \tilde{p}_{i} = i\hat{p}_{i} - \frac{i(k - 1)}{2} \hat{x}_{i}. \]
The relations for these operators can be re-expressed as in the following proposition.

\begin{proposition}
The $\hat{q}_{i}$ are symmetric and satisfy the relations
\begin{align}
[\hat{q}_{j}, \hat{x}_{i}] &= i(\delta_{ij} - \hat{x}_{i}\hat{x}_{j}) \\
[\hat{q}_{i}, \hat{q}_{j}] &= \hat{x}_{i}\hat{q}_{j} - \hat{x}_{j}\hat{q}_{i} \\
\sum _{i = 1} ^{k} \hat{x}_{i}\hat{q}_{i} &= -\frac{i(k - 1)}{2}.
\end{align}
\end{proposition}
\begin{proof}
For the first relation, we have
\begin{align*}
[\hat{q}_{j}, \hat{x}_{i}] &= i[\hat{p}_{j}, \hat{x}_{i}] + \frac{i(k - 1)}{2}[\hat{x}_{i}, \hat{x}_{j}] \\
&= i(\delta_{ij} - \hat{x}_{i}\hat{x}_{j}).
\end{align*}
For the second, we have
\begin{align*}
[\hat{q}_{i}, \hat{q}_{j}] &= -[\hat{p}_{i}, \hat{p}_{j}] - \frac{1}{2}[\hat{p}_{i}, \operatorname{div} \tilde{p}_{j}] - \frac{1}{2}[\operatorname{div} \tilde{p}_{i}, \hat{p}_{j}] \\
&= -(\hat{x}_{j}\hat{p}_{i} - \hat{x}_{i}\hat{p}_{j}) + \frac{k - 1}{2}(-(\delta_{ij} - \hat{x}_{i}\hat{x}_{j}) + (\delta_{ij} - \hat{x}_{i}\hat{x}_{j})) \\
&= \hat{x}_{i}\left(\hat{q}_{j} + \frac{i(k - 1)}{2}\hat{x}_{j}\right) - \hat{x}_{j}\left(\hat{q}_{i} + \frac{i(k - 1)}{2}\hat{x}_{i}\right) \\
&= \hat{x}_{i}\hat{q}_{j} - \hat{x}_{j}\hat{q}_{i}.
\end{align*}
For the third, we have
\begin{align*}
0 &= \sum _{i = 1} ^{k} \hat{x}_{i}\hat{p}_{i} \\
&= \sum _{i = 1} ^{k} \hat{x}_{i} \left(\hat{q}_{i} + \frac{i(k - 1)}{2}\hat{x}_{i}\right) \\
&= \sum _{i = 1} ^{k} \hat{x}_{i}\hat{q}_{i} + \frac{i(k - 1)}{2}\sum _{i = 1} ^{k} \hat{x}_{i}^{2}
\end{align*}
and rearranging gives
\[ \sum _{i = 1} ^{k} \hat{x}_{i}\hat{q}_{i} = -\frac{i(k - 1)}{2}. \qedhere \]
\end{proof}

Lastly, we re-express the kinetic energy operator in terms of the $\hat{p}_{i}$ and $\hat{q}_{i}$.
\begin{lemma}
If $f \in C^{\infty}(S^{k})$, then $\Delta f = -\sum _{i = 1} ^{k} (\hat{p}_{i})^{2}f$.
\end{lemma}
\begin{proof}
We have $\Delta f = -\operatorname{div} \nabla f$. Applying Lemma 1 gives
\begin{align*}
\operatorname{div} \nabla f &= \operatorname{div} ((\nabla f) \circ p \circ j^{-1}) \\
&= \operatorname{div} ((\nabla f) \circ (x \mapsto x/|x|))\upharpoonright_{S^{k}} \\
&= \operatorname{div} (\hat{p}_{i}f \circ (x \mapsto x/|x|))_{i \in [k + 1]} \upharpoonright_{S^{k}} \\
&= \sum _{i = 1} ^{k + 1} (\partial_{i} (\hat{p}_{i}f \circ (x \mapsto x/|x|))) \upharpoonright_{S^{k}} \\
&= \sum _{i = 1} ^{k + 1} \hat{p}_{i}^{2}f
\end{align*}
where the last line uses Lemma 2 applied to $\hat{p}_{i}f$ for each $i \in [k + 1]$.
\end{proof}

The following lemma gives the expression of the Laplacian in terms of the $\hat{q}_{i}$.

\begin{lemma}
If $f \in C^{\infty}(S^{k})$, then
\[ \Delta f = \sum _{i = 1} ^{k} \hat{q}_{i}^{2} f - \frac{(k - 1)^{2}}{4}f. \]
\end{lemma}
\begin{proof}
Using Lemma 3, we have
\begin{align*}
\Delta f &= -\sum _{i = 1} ^{k} (\hat{p}_{i})^{2} f \\
&= \sum _{i = 1} ^{k} \left(\hat{q}_{i} + \frac{i(k - 1)}{2}\hat{x}_{i}\right)^{2} f \\
&= \sum _{i = 1} ^{k} \left(\hat{q}_{i}^{2} + \frac{i(k - 1)}{2}(\hat{x}_{i}\hat{q}_{i} + \hat{q}_{i}\hat{x}_{i}) - \frac{(k - 1)^{2}}{4}\hat{x}_{i}^{2}\right) f \\
&= \sum _{i = 1} ^{k} \left(\hat{q}_{i}^{2} + \frac{i(k - 1)}{2}(2\hat{x}_{i}\hat{q}_{i} + i(1 - \hat{x}_{i}^{2})) - \frac{(k - 1)^{2}}{4}\hat{x}_{i}^{2}\right) f \\
&= \sum _{i = 1} ^{k} \hat{q}_{i}^{2} f - \frac{k(k - 1)}{2}f + \left(\frac{k - 1}{2} - \frac{(k - 1)^{2}}{4} \right)\sum _{i = 1} ^{k} \hat{x}_{i}^{2}f + i(k - 1)\sum _{i = 1} ^{k} \hat{x}_{i}\hat{q}_{i}f \\
&= \sum _{i = 1} ^{k} \hat{q}_{i}^{2} f - \frac{(k - 1)^{2}}{2}f - \frac{(k - 1)^{2}}{4}f + \frac{(k - 1)^{2}}{2}f \\
&= \sum _{i = 1} ^{k} \hat{q}_{i}^{2} f - \frac{(k - 1)^{2}}{4}f. \qedhere
\end{align*}
\end{proof}

\section{Product state ratio}
For a self-adjoint Hamiltonian $A$ acting on $H^{\otimes n}$, define
\[ h_{\operatorname{sep}}(A) \coloneqq \inf _{f_{1}, \dots, f_{n} \in P(H)} \langle f_{1} \otimes \dots \otimes f_{n}, H(f_{1} \otimes \dots \otimes f_{n}) \rangle \]
to be the optimal product-state energy. For a family $F$ of Hamiltonians, define
\[ \operatorname{PROD}(F) \coloneqq \sup _{H \in F} \frac{h_{\operatorname{sep}}(A)}{h_{0}(A)} \]
to be the optimal product-state approximation ratio.

For the rotor model, our goal is to calculate $\operatorname{PROD}(\{H_{G,n,a,b}\})$. We have
\[ \operatorname{PROD}(\{H_{G,n,a,b}\}) \geq \operatorname{PROD}(\{H_{G,2,a,b} : a, b \geq 0 \}) = \sup _{a, b \geq 0} \frac{h_{\operatorname{sep}}(H_{G,2,a,b})}{h_{0}(H_{G,2,a,b})}\] where $G$ is a graph with two vertices and a single edge between them. We will bound the numerator and denominator separately.

\subsection{Optimal product state on a single edge}
\subsubsection{Warm-up: optimal \textsc{Quantum Max-Cut} product state on a single edge}
As a warm-up, we give another proof of the following result from \cite{Gharibian_Parekh_2019}. This new proof is written in a way which will generalize more easily to the quantum rotor model and other Hamiltonians.

\begin{proposition}
The optimal product state for \textsc{Quantum Max-Cut} on a single edge has energy $\frac{1}{2}$.
\end{proposition}
\begin{proof}
Let $\psi \otimes \phi$ be a 2-qubit product state. We have
\begin{align*}
&\phantom{{}={}} \langle \psi \otimes \phi | H | \psi \otimes \phi \rangle \\
&= \frac{1 - \langle \psi \otimes \phi | X \otimes X | \psi \otimes \phi \rangle - \langle \psi \otimes \phi | Y \otimes Y | \psi \otimes \phi \rangle - \langle \psi \otimes \phi | Z \otimes Z | \psi \otimes \phi \rangle}{4} \\
&= \frac{1 - \langle \psi | X | \psi \rangle \langle \phi | X | \phi \rangle - \langle \psi | Y | \psi \rangle \langle \phi | Y | \phi \rangle - \langle \psi | Z | \psi \rangle \langle \phi | Z | \phi \rangle}{4} \\
&= \frac{1 - (\langle \psi | X | \psi \rangle, \langle \psi | Y | \psi \rangle, \langle \psi | Z | \psi \rangle) \cdot (\langle \phi | X | \phi \rangle, \langle \phi | Y | \phi \rangle, \langle \phi | Z | \phi \rangle)}{4}.
\end{align*}
The two vectors $(\langle \psi | X | \psi \rangle, \langle \psi | Y | \psi \rangle, \langle \psi | Z | \psi \rangle)$ and $(\langle \phi | X | \phi \rangle, \langle \phi | Y | \phi \rangle, \langle \phi | Z | \phi \rangle)$ are just the points on the Bloch sphere corresponding to $\psi$ and $\phi$ respectively. Thus their inner product is at most 1, which implies
\[  \langle \psi \otimes \phi | H | \psi \otimes \phi \rangle \leq \frac{1 - (-1)}{4} = \frac{1}{2}. \qedhere \]
\end{proof}

The fact that for a one-qubit state $\psi$ the vector of expectation values $(\langle X \rangle_{\psi}, \langle Y \rangle_{\psi}, \langle  Z \rangle_{\psi})$ lies on the Bloch sphere is a particular instance of an uncertainty principle (analogous to the Heisenberg uncertainty principle) for qubit states. The corresponding step for the rotor model will use an uncertainty principle for wavefunctions on a sphere due to Erb \cite{Erb_2010}.

\subsubsection{Optimal rotor product state on a single edge}
For the numerator, we have
\[ H_{G,2,a,b} \coloneqq a(\Delta_{1} + \Delta_{2}) + b(1 + x_{1} \cdot x_{2}). \]
and the energy of a given product state is
\[ \langle f \otimes g, H(f \otimes g) \rangle = a(\langle f, \Delta f \rangle + \langle g, \Delta g \rangle) + b\left(1 + \sum _{i = 1} ^{k} \langle f, x_{i} f \rangle \langle g, x_{i} g \rangle\right). \]
In particular, the energy depends only on the expectation values $\langle f, \Delta f \rangle$ and $\langle f, x_{i} f \rangle$ and likewise for $g$. Tight sets of constraints on such expectation values are usually referred to as uncertainty principles. For the sphere, a tight uncertainty principle seems to have been rediscovered multiple times in the literature \cite{Erb_2010, Goodman_Goh_2004, Ogawa_Nagasawa_2022, Rosler_Voit_1997}. The most general one seems to be due to \cite[Corollary 2.60]{Erb_2010}, and it is
\[ \langle f, \Delta_{S^{k - 1}} f \rangle \geq \left(\frac{k - 1}{2}\right)^{2} \frac{|\mu(f)|^{2}}{1 - |\mu(f)|^{2}} \]
where $\mu(f) \coloneqq (\langle f, x_{i} f \rangle)_{i \in [k]}$ is the spherical mean of $f$ and $f$ is normalized such that $|f|_{L^{2}(S^{k})} = 1$.

If we choose $f, g$ saturating the uncertainty principle up to a universal constant $C$ such that $\mu(f) = se_{1}$, $\mu(g) = -te_{1}$, this shows
\[ h_{\operatorname{sep}}(H_{G,2,a,b}) = \inf _{(s, t) \in [0, 1]^{2}} Ca\left(\frac{k - 1}{2}\right)^{2}\left(\frac{t^{2}}{1 - t^{2}} + \frac{s^{2}}{1 - s^{2}}\right) + b(1 - st). \]

\subsection{Optimal entangled state on a single edge}
To bound $h_{0}(H_{G,2,a,b})$, let $f_{x,t}$ saturate the uncertainty principle up to a universal constant C with $\mu(f_{x, t}) = tx$ for $t \in [0, 1]$ and $x \in S^{k - 1}$. Consider the family of states
\begin{align*}
\psi(x_{1}, x_{2}) &\coloneqq f_{-x_{2},t}(x_{1}) \\
&= f_{-x_{1},t}(x_{2})
\end{align*}
where equality holds since $f_{x,t}$ is radial and $f_{x,t}(y)$ depends only on $x \cdot y$. Then
\begin{align*}
h_{0}(H_{G,2,a,b}) &\leq \langle \psi, H \psi \rangle \\
&= a\int _{S^{k-1}} \langle f_{-x_{2},t}, \Delta f_{-x_{2},t} \rangle \,d\nu(x_{2}) + a\int_{S^{k-1}} \langle f_{-x_{1},t}, \Delta f_{-x_{1},t} \rangle \,d\nu(x_{1}) \\
&\phantom{{}={}} + b\int _{S^{k - 1}} \int _{S^{k - 1}} (1 + x_{1} \cdot x_{2}) |f_{-x_{2},t}(x_{1})|^{2} \,d\nu(x_{1})\,d\nu(x_{2}) \\
&= 2aC \left(\frac{k - 1}{2}\right)^{2}\frac{t^{2}}{1 - t^{2}} + b\left(1 + \int _{S^{k - 1}} x_{2} \cdot \int _{S^{k - 1}} x_{1} |f_{-x_{2},t}(x_{1})|^{2}\,d\nu(x_{1})\,d\nu(x_{2})\right) \\
&= 2aC \left(\frac{k - 1}{2}\right)^{2}\frac{t^{2}}{1 - t^{2}} + b\left(1 + \int _{S^{k - 1}} x_{2} \cdot (-tx_{2})\,d\nu(x_{2})\right) \\
&= 2aC \left(\frac{k - 1}{2}\right)^{2}\frac{t^{2}}{1 - t^{2}} + b(1 -t).
\end{align*}

\subsection{Ratio}
Combining the two arguments shows
\begin{align*}
\operatorname{PROD}(\{H_{G,n,a,b}\}) &\geq \sup _{a, b \geq 0} \frac{\inf _{(s, t) \in [0, 1]^{2}} a(\frac{k - 1}{2})^{2}(\frac{t^{2}}{1 - t^{2}} + \frac{s^{2}}{1 - s^{2}}) + b(1 - st)}{\inf _{t \in [0, 1]} 2a \left(\frac{k - 1}{2}\right)^{2}\frac{t^{2}}{1 - t^{2}} + b(1 -t)} \\
&= \sup _{a, b \geq 0} \frac{\inf _{(s, t) \in [0, 1]^{2}} a(\frac{t^{2}}{1 - t^{2}} + \frac{s^{2}}{1 - s^{2}}) + b(1 - st)}{\inf _{t \in [0, 1]} 2aC \frac{t^{2}}{1 - t^{2}} + b(1 -t)}.
\end{align*}
The two suprema are equal, since $a$ can be rescaled by $(\frac{k - 1}{2})^{2}$. In particular, this shows that the optimal product state approximation ratio is bounded below by a universal constant which is independent of $k$.

\subsection{Sum-of-squares proof of spherical uncertainty principle}\label{sec:spherical-uncertainty}
\subsubsection{Warm-up: SoS proof of Bloch sphere uncertainty principle}
\begin{proposition}
Let $\psi \in \mathbb{C}^{2}$ be a qubit state. Then $\langle X \rangle_{\psi}^{2} + \langle Y \rangle_{\psi}^{2} + \langle Z \rangle_{\psi}^{2} \leq 1$.
\end{proposition}
\begin{proof}
Consider the moment matrix of $\psi$ with respect to the observables $1, X, Z$. It is
\[ M \coloneqq \begin{bmatrix}
1 & \langle X \rangle_{\psi} & \langle Z \rangle_{\psi} \\
\langle X \rangle_{\psi} & 1 & i\langle Y \rangle_{\psi} \\
\langle Z \rangle_{\psi} & -i\langle Y \rangle_{\psi} & 1
\end{bmatrix}. \]
Since $M$ is PSD, we have
\[ 0 \leq \det M = 1 - \langle X \rangle_{\psi}^{2} - \langle Y \rangle_{\psi}^{2} - \langle Z \rangle_{\psi}^{2}. \qedhere \]
\end{proof}

\subsubsection{Second warm-up: SoS proof of the Heisenberg uncertainty principle}
\begin{proposition}
Let $f \in C^{\infty}(\mathbb{R}) \cap L^{2}(\mathbb{R})$ be a continuous-variable state. Then
\[ (\langle x^{2} \rangle_{f} - \langle x \rangle^{2}_{f})(\langle p^{2} \rangle_{f} - \langle p \rangle^{2}_{f}) \geq \hbar^{2} / 4. \]
\end{proposition}
\begin{proof}[Proof (\cite{Curtright_Zachos_2001})]
Consider the moment matrix of $f$ with respect to the observables $1, x, p$. It is
\[ M \coloneqq \begin{bmatrix}
1 & \langle x \rangle_{f} & \langle p \rangle_{f} \\
\langle x \rangle_{f} & \langle x^{2} \rangle_{f} & \langle xp \rangle_{f} \\
\langle p \rangle_{f} & \langle px \rangle_{f} & \langle p^{2} \rangle_{f}
\end{bmatrix}. \]
The Schur complement of $M$ with respect to the upper-left 1-by-1 block of $1$ is
\[ M/1 = \begin{bmatrix}
\langle x^{2} \rangle_{f} - \langle x \rangle_{f}^{2} & \langle xp \rangle_{f} - \langle x \rangle_{f} \langle p \rangle_{f} \\
\langle px \rangle_{f} - \langle p \rangle_{f} \langle x \rangle_{f} & \langle p^{2} \rangle_{f} - \langle p \rangle_{f}^{2} 
\end{bmatrix} \]
and must be PSD since both $M$ and $[1]$ are PSD. Thus
\begin{align*}
0 &\leq \det(M/1) \\
&= (\langle x^{2} \rangle_{f} - \langle x \rangle_{f}^{2})(\langle p^{2} \rangle_{f} - \langle p \rangle_{f}^{2}) - (\langle xp \rangle_{f} - \langle x \rangle_{f} \langle p \rangle_{f})(\langle px \rangle_{f} - \langle x \rangle_{f} \langle p \rangle_{f}) \\
&= (\langle x^{2} \rangle_{f} - \langle x \rangle_{f}^{2})(\langle p^{2} \rangle_{f} - \langle p \rangle_{f}^{2}) - (\Re \langle xp \rangle_{f} - \langle x \rangle_{f} \langle p \rangle_{f})^{2} - \hbar^{2}/4
\end{align*}
where in the last line we have used the canonical commutation relation $[x, p] = i\hbar$ and the fact that the moment matrix must respect it.
\end{proof}

\subsubsection{SoS proof of the spherical uncertainty principle}
We now give an ncSoS proof of the spherical uncertainty principle, which uses a slightly different set of momentum operators on the sphere from the operators used in the SDP relaxation. This proof can be rewritten as a degree-4 ncSoS proof with respect the operators used in our SDP.

\begin{proposition}
Let $f \in C^{\infty}(S^{k - 1})$. Then
\[ \langle f, \Delta_{S^{k-1}} f \rangle \geq \left( \frac{k - 1}{2} \right)^{2} \frac{|\mu(f)|^{2}}{1 - |\mu(f)|^{2}} \]
where $\mu(f) \coloneqq (\langle f, x_{i} f\rangle)_{i \in [k]}$ is the spherical mean of $f$, the norm of $\mu(f)$ is taken to be the $L^{2}$ norm in $\mathbb{R}^{k}$, and $f$ is normalized such that $|f|_{L^{2}(S^{k})} = 1$.
\end{proposition}
\begin{proof}
 We can assume without loss of generality that $\mu(f) = te_{1}$ by composing $f$ with a suitable rotation. Consider the vector fields $v_{ij} = x_{i}e_{j} - x_{j}e_{i}$ on $S^{k - 1}$ for $1 \leq i < j \leq k$. We have
\[ [v_{ij}, x_{k}] = \begin{cases}
ix_{i} & k = j \\
-ix_{j} & k = i \\
0 & k \neq i, j
\end{cases}. \]
A standard calculation shows that the $v_{ij}$ are divergence free and that
\[ \Delta_{S^{k - 1}} = -\sum _{1 \leq i < j \leq k} v_{ij}^{2}. \]
Let $M$ be the moment matrix of $f$ with respect to the operators $1, (x_{i})_{i \in [k]}, (v_{ij})_{\{i, j\} \in \binom{[k]}{2}}$. Consider the action of $O(k - 1)$, which acts on $\operatorname{span} \{x_{i}\}$ as a copy of $1 \oplus \text{std}$ and on $\operatorname{span} \{v_{ij}\}$ as a copy of $\wedge^{2} (1 \oplus \text{std})$. If we replace $M$ by its invariant part under this action then this does not change $M[x_{i}]$ or $M[\Delta_{S^{k - 1}}]$, so we can assume that $M$ is invariant.

We have
\[ \wedge^{2} (1 \oplus \text{std}) \cong \text{std} \oplus \wedge^{2} \text{std} \]
so by Schur's lemma
\[ M = \begin{bmatrix}
1 & t \\
t & M[x_{1}^{2}]
\end{bmatrix} \oplus \left(\begin{bmatrix}
M[x_{2}^{2}] & M[x_{2}v_{12}] \\
M[v_{12}x_{2}] & M[v_{12}^{2}]
\end{bmatrix} \otimes I_{k - 1}\right) \oplus M[v_{23}^{2}]I_{\binom{k - 1}{2}}. \]
Since $M$ is PSD, we have $M[x_{1}^{2}] \geq t^{2}$ and $\sum _{i = 1} ^{k} M[x_{i}^{2}] = 1$, so $\sum _{i = 2} ^{k} M[x_{i}^{2}] \leq 1 - t^{2}$. Also, $\Im M[x_{i}v_{1i}] = M[x_{i}]/2 = t/2$. Taking the partial trace of the second block implies that
\[ \begin{bmatrix}
\sum _{i = 2} ^{k} M[x_{i}^{2}] & \sum _{i = 2} ^{k} M[x_{i}v_{1i}] \\
\sum _{i = 2} ^{k} M[x_{i}v_{1i}] & \sum _{i = 2} ^{k} M[v_{1i}^{2}]
\end{bmatrix} \]
is PSD, so taking its determinant gives
\begin{align*}
0 &\leq \left( \sum _{i = 2} ^{k} M[v_{1i}^{2}] \right) \left( \sum _{i = 2} ^{k} M[x_{i}^{2}] \right) - \left( \left( \sum _{i = 2} ^{k} \Re M[x_{i}v_{1i}] \right)^{2} + \left( \frac{(k - 1)t}{2} \right)^{2} \right) \\
&\leq \left( \sum _{i = 2} ^{k} M[v_{1i}^{2}] \right) (1 - t^{2}) - \left( \frac{k - 1}{2} \right)^{2} t^{2}.
\end{align*}
Thus
\[ \langle f, \Delta_{S^{k - 1}} f\rangle = \sum _{1 \leq i < j \leq k} M[v_{ij}^{2}] \geq \sum _{i = 2} ^{k} M[v_{1i}^{2}] \geq \left( \frac{k - 1}{2} \right)^{2} \frac{t^{2}}{1 - t^{2}}. \qedhere \]
\end{proof}

\section{SDP rounding and approximation ratio}
\subsection{SDP} We use an SDP based on level 1 of the ncSoS hierarchy, with respect to the variables
\[ \cup \{\{x_{v, i}, \hat{q}_{v,i}\} : v \in V(G), i \in [k + 1]\}. \]
The SDP optimizes over pseudo-expectations defined on all noncommutative polynomials in these variables with degree at most 2. A valid degree-2 pseudo-expectation $\tilde{E}$ must satisfy the constraints
\begin{align*}
\tilde{E}[x_{v,i}x_{w,j}] &= \tilde{E}[x_{w,j}x_{v,i}] \\
\tilde{E}[x_{v,i}\hat{q}_{w,j}] &= \tilde{E}[\hat{q}_{w,j}x_{v,i}] \\
\tilde{E}[\hat{q}_{v,i}\hat{q}_{w,j}] &= \tilde{E}[\hat{q}_{w,j}\hat{q}_{v,i}]
\end{align*}
for $v \neq w$, and
\begin{align*}
\tilde{E}[x_{v,i}x_{v,j}] &= \tilde{E}[x_{v,j}x_{v,i}] \\
\tilde{E}[x_{v,i}\hat{q}_{v,j}]^{*} &= \tilde{E}[\hat{q}_{v,j}x_{v,i}] \\
\tilde{E}[\hat{q}_{v,i}\hat{q}_{v,j}]^{*} &= \tilde{E}[\hat{q}_{v,j}\hat{q}_{v,i}] \\
\Im \tilde{E}[x_{v,i}\hat{q}_{v,j}] &= \frac{i}{2}(\delta_{ij} - \tilde{E}[x_{v,i}x_{v,j}]) \\
\Im \tilde{E}[\hat{q}_{v,i}\hat{q}_{w,j}] &= \frac{i}{2}(\tilde{E}[x_{v,i}\hat{q}_{v,j} - x_{v,j}\hat{q}_{v,i}]) \\
\sum _{i = 1} ^{k} \tilde{E}[x_{v,i}^{2}] &= 1 \\
\sum _{i = 1} ^{k} \tilde{E}[x_{v,i}q_{v,i}] &= -\frac{i(k - 1)}{2}
\end{align*}
for all $v \in V(G)$.

Additionally, a valid pseudo-expectation must satisfy a positivity constraint. This constraint will be equivalent to the moment matrix being PSD. We will write the moment matrix $M$ by splitting it into blocks corresponding to each site, taking into account the commutation relations of the $\hat{x}_{i}$ and $\hat{q}_{j}$ and the fact that $\tilde{E}$ must satisfy them. More precisely, we define
\begin{align*}
M_{1,1} &= [1] \\
M_{1,v} &= [\tilde{E}(x_{v,i})]_{i \in [k]} \oplus [\tilde{E}(\hat{q}_{v,i})]_{i \in [k]} \\
M_{v,w} &= \begin{bmatrix}
[\tilde{E}(x_{v,i}x_{w,j})]_{i,j \in [k]} & [\tilde{E}(x_{v,i}\hat{q}_{w,j})]_{i,j \in [k]} \\
[\tilde{E}(\hat{q}_{v,j}x_{w,i})]_{i,j \in [k]} & [\tilde{E}(\hat{q}_{v,j}\hat{q}_{w,i})]_{i,j \in [k]}
\end{bmatrix}
\end{align*}
and the constraint on the overall moment matrix $M$ is that $M \succeq 0$.

The objective function of the SDP is
\[ a \sum _{v \in V(G)}  \left( -\frac{(k - 1)^{2}}{4} + \sum _{i = 1} ^{k} \tilde{E}[\hat{q}_{v,i}^{2}] \right) + b \sum _{(v, w) \in E(G)} \left(1 + \sum _{i = 1} ^{k} \tilde{E}[x_{v,i}x_{w,i}]\right). \]

\subsection{Symmetry reduction of the SDP}\label{sec:symmetry}
The Hamiltonian has a global on-site $O(k)$ symmetry, given by simultaneously rotating the coordinates of each rotor by the same orthogonal matrix. We will show that this induces a symmetry of the SDP and thus that we can assume the optimal solution is invariant under the $O(k)$ action. Similarly to the SDP from \cite{Briet_Oliveira_Vallentin_2014}, this will also show that the SoS-based SDP is equivalent to another SDP whose size depends only on $n = \#V(G)$ and not on $k$.

We define an action of $O(k)$ on the free algebra generated by $\hat{x}_{i}$ and $\hat{q}_{j}$ by
\begin{align*}
Q \cdot \hat{x}_{i} &= \sum _{j = 1} ^{k} Q_{j,i}\hat{x}_{j} \\
Q \cdot \hat{q}_{i} &= \sum _{j = 1} ^{k} Q_{j,i}\hat{q}_{j}
\end{align*}
and show in the next proposition that this gives a well-defined action after taking the quotient by the relations defined in Section 3.

\begin{proposition}
The two-sided ideal of relations among the $\hat{x}_{i}$ and $\hat{q}_{j}$ deduced in Section 3 is $O(k)$-invariant.
\end{proposition}
\begin{proof}
For convenience, define
\begin{align*}
\hat{x}_{u} &\coloneqq \sum _{i = 1} ^{k} u_{i}\hat{x}_{i} \\
\hat{q}_{u} &\coloneqq \sum _{i = 1} ^{k} u_{i}\hat{q}_{i}
\end{align*}
for any vector $u \in \mathbb{R}^{k}$.
We have
\begin{align*}
[\hat{x}_{u}, \hat{q}_{v}] &= i\sum _{i = 1} ^{k} \sum _{j = 1} ^{k} u_{i}v_{j}(\delta_{ij} - \hat{x}_{i}\hat{x}_{j}) \\
&= i(u \cdot v) - i\sum _{i = 1} ^{k} \sum _{j = 1} ^{k} (u_{i}\hat{x}_{i})(v_{j}\hat{v}_{j}) \\
&= i(u \cdot v - \hat{x}_{u}\hat{x}_{v})
\end{align*}
and
\begin{align*}
[\hat{q}_{u}, \hat{q}_{v}] &= \sum _{i = 1} ^{k} \sum _{j = 1} ^{k} u_{i}v_{j}(\hat{x}_{i}\hat{q}_{j} - \hat{x}_{j}\hat{q}_{i}) \\
&= \sum _{i = 1} ^{k} \sum _{j = 1} ^{k} ((u_{i}\hat{x}_{i})(v_{j}\hat{q}_{j}) - (v_{j}\hat{x}_{j})(u_{i}\hat{q}_{i})) \\
&= \hat{x}_{u}\hat{q}_{v} - \hat{x}_{v}\hat{q}_{u}.
\end{align*}
If $u_{1}, \dots, u_{k}$ are the columns of an orthogonal matrix $Q$ then
\begin{align*}
\sum _{i = 1} ^{k} \hat{x}_{u_{i}}^{2} &= \sum _{i = 1} ^{k} \sum _{j = 1} ^{k} \sum _{l = 1} ^{k} u_{i,j}u_{i,l}\hat{x}_{j}\hat{x}_{l} \\
&= \sum _{j = 1} ^{k} \sum _{l = 1} ^{k} \hat{x}_{j}\hat{x}_{l} \sum _{i = 1} ^{k} u_{i,j}u_{i,l} \\
&= \sum _{j = 1} ^{k} \sum _{l = 1} ^{k} \hat{x}_{j}\hat{x}_{l} \delta_{j,l} \\
&= \sum _{i = 1} ^{k} \hat{x}_{i}^{2}
\end{align*}
where the third line follows since the rows of $Q$ are orthogonal, and
\begin{align*}
\sum _{i = 1} ^{k} \hat{x}_{u_{i}}\hat{q}_{u_{i}} &= \sum _{i = 1} ^{k} \sum _{j = 1} ^{k} \sum _{l = 1} ^{k} u_{i,j}u_{i,l}\hat{x}_{j}\hat{q}_{l} \\
&= \sum _{j = 1} ^{k} \sum _{l = 1} ^{k} \hat{x}_{j}\hat{q}_{l} \sum _{i = 1} ^{k} u_{i,j}u_{i,l} \\
&= \sum _{j = 1} ^{k} \sum _{l = 1} ^{k} \hat{x}_{j}\hat{q}_{l} \delta_{j,l} \\
&= \sum _{j = 1} ^{k} \hat{x}_{j}\hat{q}_{j}
\end{align*}
where again the third line follows since the rows of $Q$ are orthogonal.
\end{proof}

The action of $O(k)$ on the algebra generated by the $\hat{x}_{i}$ and $\hat{q}_{j}$ induces an action on pseudo-expectations $\tilde{E}$, by
\[ (Q \cdot \tilde{E})(f) = \tilde{E}[Q^{-1} \cdot f] \]
for any noncommutative polynomial $f$. On moment matrices, the action is then given by the formulas
\begin{align*}
(Q \cdot M)_{1,1} &= [1] \\
(Q \cdot M)_{1,v} &= (Q^{T} \oplus Q^{T})M_{1,v} \\
(Q \cdot M)_{v,w} &= (Q^{T} \oplus Q^{T})M_{v,w}(Q \oplus Q).
\end{align*}

\begin{lemma}
The SDP in Section 4.2.1 is $O(k)$-invariant.
\end{lemma}

\begin{proposition}
There is a solution to the SDP whose moment matrix $M$ is of the form
\[ M = 1 \oplus M' \otimes I_{k} \]
for some $M' \in \mathbb{C}^{2n \times 2n}$ and where $n = \#V(G)$.
\end{proposition}

The constraints on the SDP will force $M'$ to satisfy certain constraints. In particular, $M'$ must be of the form
\begin{align*}
M'_{v,v} &= \begin{bmatrix}
1/k & K_{v,v} + \frac{i(k - 1)}{2k} \\
K_{v,v} - \frac{i(k - 1)}{2k} & L_{v,v}
\end{bmatrix} \\
M'_{v, w} &\in \mathbb{R}^{2 \times 2}
\end{align*}
for some $L_{v,v} \geq 0$ and $K_{v,v} \in \mathbb{R}$.

\section{Rounding and analysis of the approximation ratio}\label{sec:rounding-analysis}
The main idea in this paper is a strategy for ``rounding'' the output (a ``pseudo-state'') of the ncSoS SDP to a genuine quantum state. The motivation comes from the general idea of a duality between algebra and geometry: any ``geometric space'' $X$ is dual to some algebra $A(X)$ of functions $X \to \mathbb{C}$, with structure on the space corresponding to structure on the algebra, and a map $X \to Y$ between two spaces corresponds to a homomorphism $A(Y) \to A(X)$ given by precomposition. Traditionally, the most common examples of this are Gelfand duality between compact Hausdorff topological spaces and $C^{*}$-algebras, and Zariski duality between varieties (affine schemes) and reduced commutative rings (arbitrary commutative rings). There has also been further work on other kinds of spaces and algebras, such as for smooth manifolds or for measurable spaces and von Neumann algebras.

The reason for considering this duality is that in quantum mechanics, the notion of geometric space with an underlying set of points often fails (because of the uncertainty principle), but the algebra of operators always exists and is generally some noncommutative ring. In fact, the Heisenberg picture of quantum mechanics is based entirely on the algebra of observables and is a common formalism in physics. Connes later introduced the idea of studying ``non-commutative geometry'' using non-commutative operator algebras. However, the Heisenberg picture is dual to the more common Schr\"{o}dinger picture, under which an observable $A$ in a non-commutative algebra corresponds to the expectation functional $\rho \mapsto \langle A \rangle_{\rho}$ mapping a state $\rho$ to the expectation value of $A$ under $\rho$. In general, $\rho$ exists as a quantum state but cannot be localized to a classical point without violating an uncertainty principle. A homomorphism between the algebras of observables then becomes a quantum channel between the spaces of density operators.

\subsection{Classical rounding in the Heisenberg picture}
Classically, the rounding map for the rank-$k$ Grothendieck problem was defined as a map $\mathbb{R}^{k} \to S^{k - 1}$
\[ f(x) = \begin{cases} x/|x| & x \neq 0 \\
\text{arbitrary} & x = 0 \end{cases} \]
In the ``classical Heisenberg picture,'' the dual is a homomorphism $L^{\infty}(S^{k - 1}) \to L^{\infty}(\mathbb{R}^{k})$ of von Neumann algebras given by
\[ g \mapsto g \circ f \]
where $g : \mathbb{R}^{k} \to \mathbb{C}$. Taking the dual again to go to the Schr\"{o}dinger picture, the corresponding classical channel is a Markov kernel
\[ \mathcal{P}(\mathbb{R}^{k}) \cong \mathcal{P}(\mathbb{R}_{> 0} \times S^{k - 1}) \to \mathcal{P}(S^{k - 1}) \]
which takes a probability distribution over $\mathbb{R}^{k}$, reparameterizes it into spherical coordinates, and then marginalizes out the radial part of the distribution to get probability distribution over $S^{k - 1}$.

\subsection{Defining the rounding map}
For convenience, redefine $M = M' \otimes I_{k}$ where $M'$ is as in Proposition 8. The algebra used for the rounding will be (roughly) the algebra of operators on $L^{2}(\mathbb{R}^{n \times k})$ and the state will chosen as a bosonic Gaussian state. (Since the Laplacian is unbounded it is not entirely clear what kind of operator algebra to choose, so we do not yet claim that the rounding map can be specified as a map between operator algebras.)

In the Schr\"{o}dinger picture, the rounding map on a given site will be
\[ \operatorname{tr}_{2} : B(L^{2}(\mathbb{R}^{k})) \cong B(L^{2}(S^{k - 1}) \otimes L^{2}(\mathbb{R}_{>0}, r^{k-1}\,dr)) \to B(L^{2}(S^{k - 1})). \]
That is, to obtain the rounded state given a density operator $\rho$ on $L^{2}(\mathbb{R}^{k})$, first use the standard isomorphism $L^{2}(\mathbb{R}^{k}) \cong L^{2}(S^{k - 1}) \otimes L^{2}(\mathbb{R}_{>0}, r^{k-1}\,dr)$ (given by the transformation between Cartesian and spherical coordinates) to get a density operator on $L^{2}(S^{k - 1}) \otimes L^{2}(\mathbb{R}_{>0}, r^{k-1}\,dr)$, and then trace out the second tensor factor to get a density operator on $L^{2}(S^{k - 1})$.

To calculate the rounding map in the Heisenberg picture, let $\psi \in L^{2}(\mathbb{R}^{k})$ be a pure state and $A \in B(L^{2}(S^{k - 1}))$ be a self-adjoint (possibly unbounded) operator such that $\langle \psi, A \psi \rangle$ exists and is finite. 
The density operator of $\psi$ is the rank-1 projection
\[ \rho_{\psi}(f) \coloneqq \psi \langle \psi, f \rangle. \]
Let $\psi = \sum _{i} f_{i} \otimes g_{i}$ be the Schmidt decomposition of $\psi$ after doing the spherical change of coordinates.
The expectation value of $A$ on the rounded state is
\begin{align*}
\operatorname{tr}(A\operatorname{tr}_{2}(\rho_{\psi})) &= \operatorname{tr}((A \otimes I_{L^{2}(\mathbb{R}_{>0}, r^{k - 1}\,dr)})\rho_{\psi}) \\
&= \langle \psi, (A \otimes I_{L^{2}(\mathbb{R}_{>0}, r^{k - 1}\,dr)})\psi \rangle \\
&= \sum _{i, j} \langle f_{i} \otimes g_{i}, (Af_{j})\otimes g_{j} \rangle \\
&= \sum _{i, j} \langle f_{i}, Af_{j} \rangle\langle g_{i}, g_{j} \rangle \\
&= \sum_{i,j} \langle f_{i}, Af_{j} \rangle \int _{\mathbb{R}_{>0}} g_{i}(r)g_{j}(r) r^{k - 1} \,dr  \\
&= \int _{\mathbb{R}_{>0}} \sum_{i,j} g_{i}(r)g_{j}(r) \langle f_{i}, Af_{j} \rangle r^{k - 1} \,dr \\
&= \int _{\mathbb{R}_{>0}} \langle \psi(-, r), A\psi(-, r) \rangle \,r^{k - 1}\, dr.
\end{align*}
All of the integrals and sums in the above calculation converge absolutely and can thus be freely interchanged, since $\langle \psi, A \psi \rangle$ was assumed to be finite.



\subsection{Rounding map on Hamiltonian terms}
We now calculate the action of the rounding map on the terms in the Hamiltonian. For the position operators, we have
\begin{align*}
\operatorname{tr}(x_{i} \operatorname{tr}_{2}(\rho_{\psi})) &= \int _{\mathbb{R}_{>0}} \langle \psi(-, r), x_{i} \psi(-, r)\rangle r^{k - 1}\,dr \\
&= \int _{\mathbb{R}_{>0}} \left(\int _{S^{k-1}} \psi(x, r)^{*}x_{i}\psi(x, r)\, d\mu(x)\right)r^{k - 1}\,dr \\
&= \int _{\mathbb{R}^{k}} \psi(x)^{*} \frac{x_{i}}{|x|} \psi(x)\,dx \\
&= \left\langle \psi, \frac{x_{i}}{|x|}\psi \right\rangle.
\end{align*}

\begin{lemma}
Let $v$ be a vector field on $\mathbb{R}^{k}$ such that $v(x) \cdot x = 0$ for all $x$ and $v(rx) = rv(x)$ for all $r \geq 0$. Then $\partial_{v} f(x, r) = (\partial_{v} f)(rx)$ for all $x, r$.
\end{lemma}
\begin{proof}
We have
\begin{align*}
(\partial_{v} f)(rx) &= v(rx) \cdot \nabla f(rx) \\
&= rv(x) \cdot \nabla f(rx).
\end{align*}
and
For the spherical gradient, fix $r > 0$ and define $g(x) = f(rx)$. Then
\begin{align*}
(\nabla f(-, r))(x) &= (\nabla g(-, 1))(x) \\
&= \nabla g(x) - x(x \cdot \nabla g(x)) \\
&= r \nabla f(x) - x(x \cdot r \nabla f(x))
\end{align*}
so
\begin{align*}
\partial_{v}f(x, r) &= v(x) \cdot (\nabla f(-, r))(x) \\
&= v(x) \cdot (r \nabla f(x) - x(x \cdot r \nabla f(x))) \\
&= v(x) \cdot (r \nabla f(x)) \\
&= rv(x) \cdot \nabla f(x)
\end{align*}
where the third line follows since $v(x) \cdot x = 0$.
\end{proof}

For the Laplacian, we will write it in terms of vector fields $v_{ij}(x) = x_{i}e_{j} - x_{j}e_{i}$. Clearly $v_{ij}(x) \cdot x = 0$, so the restrictions define vector fields on $S^{k - 1}$. A standard calculation shows that the $v_{ij}$ are divergence-free and that $\Delta = \sum v_{ij}^{2}$. If $\phi \in C^{\infty}(\mathbb{R}^{k})$ we have
\begin{align*}
\operatorname{tr}(\Delta \operatorname{tr}_{2}(\rho_{\psi})) &= \int _{\mathbb{R}_{>0}} \langle \psi(-, r), \Delta \psi(-, r)\rangle r^{k - 1}\,dr \\
&= \int _{\mathbb{R}_{>0}} \sum _{1 \leq i < j \leq k} \langle \psi(-, r), v_{ij}^{2} \psi(-, r)\rangle r^{k - 1}\,dr \\
&= \int _{\mathbb{R}_{>0}} \sum _{1 \leq i < j \leq k} \langle v_{ij} \psi(-, r), v_{ij} \psi(-, r)\rangle r^{k - 1}\,dr \\
&= \int _{\mathbb{R}_{>0}} \sum _{1 \leq i < j \leq k} \left( \int _{S^{k-1}} |v_{ij} \psi(-, r)|(x)^{2}\,d\mu(x)\right) r^{k - 1}\,dr \\
&= \sum _{1 \leq i < j \leq k} \int _{\mathbb{R}_{>0}} \left( \int _{S^{k-1}} |v_{ij} \psi(-, r)|(x)^{2}\,d\mu(x)\right) r^{k - 1}\,dr \\
&= \sum _{1 \leq i < j \leq k} \int _{\mathbb{R}^{k}} |v_{ij}\psi(x)|^{2}\,dx \\
&= \int _{\mathbb{R}^{k}} \psi(x)^{*} \left(\sum _{1 \leq i < j \leq k} v_{ij}^{2}\psi(x)\right)\,dx
\end{align*}
where the sixth line uses Lemma 9.

\subsection{Weyl transform of rounded operators}
Since the $x_{v,i}$ all commute with each other, the Weyl transform of any rounded potential term is a function of the rounded position operators, that is
\[ \sum _{i = 1} ^{k} \frac{x_{v,i}x_{w,i}}{(\sum _{i = 1} ^{k} x_{v,i}^{2})^{1/2} (\sum _{i = 1} ^{k} x_{w,i}^{2})^{1/2}}. \]

The Weyl transform of the operator $v_{ij}^{2}$ can be computed using the Moyal star-product as
\begin{align*}
&\phantom{{}={}} (x_{i} * p_{j} - x_{j} * p_{i}) * (x_{i} * p_{j} - x_{j} * p_{i}) \\
&= (x_{i}p_{j} - x_{j}p_{i})*(x_{i}p_{j} - x_{j}p_{i}) \\
&= x_{i}^{2}p_{j}^{2} + x_{j}^{2}p_{i}^{2} - \left(x_{i}x_{j}p_{i}p_{j} + \frac{i\hbar}{2}(x_{i}p_{i} - x_{j}p_{j}) - \frac{1}{8}2\hbar^{2}\right) \\
&\phantom{{}={}} - \left(x_{i}x_{j}p_{i}p_{j} + \frac{i\hbar}{2}(x_{j}p_{j} - x_{i}p_{i}) - \frac{1}{8}2\hbar^{2}\right) \\
&= x_{i}^{2}p_{j}^{2} + x_{j}^{2}p_{i}^{2} - 2x_{i}x_{j}p_{i}p_{j} - \frac{1}{2}\hbar^{2} \\
&= (x_{i}p_{j} - x_{j}p_{i})^{2} - \frac{1}{2}\hbar^{2}.
\end{align*}

\subsection{Value of the rounded solution}
Let $\rho$ be the original Gaussian and $\rho'$ be the rounded state. The Wigner function $W$ of $\rho$ is a Gaussian with mean 0 and covariance $\Re \Sigma$. Thus the expectation value with respect to $\rho'$ of the potential term corresponding to an edge $(v, w)$ is $f_{BOV}(k\Sigma_{(v, 1),(w,1)})$, and the corresponding approximation ratio is $\alpha_{BOV, k}^{-1}$ from the rounding algorithm of \cite{Briet_Oliveira_Vallentin_2014}.

We now calculate the expectation value of a kinetic energy term for a vertex $v$. From the symmetry reduction argument we know $\Re \Sigma_{v,v}$ is diagonal, so the Wigner function is an independent multivariate Gaussian density. Thus
\begin{align*}
&\phantom{{}={}} \sum _{1 \leq i < j \leq k} E_{W}\left[(x_{v,i}p_{v,j} - x_{v,j}p_{v,i})^{2} - \frac{1}{2}\right] \\
&= -\frac{k(k - 1)}{4} + \sum _{1 \leq i < j \leq k} E_{W}[x_{v,i}^{2}p_{v,j}^{2} + x_{v,j}^{2}p_{v,i}^{2} - 2x_{v,i}x_{v,j}p_{v,i}p_{v,j}] \\
&= -\frac{k(k - 1)}{4} + \sum _{1 \leq i < j \leq k} E_{W}[x_{v,i}^{2}]E_{W}[p_{v,j}^{2}] + E_{W}[x_{v,j}^{2}]E_{W}[p_{v,i}^{2}] \\
&= -\frac{k(k - 1)}{4} + \sum _{1 \leq i < j \leq k} \frac{k^{2}}{(k - 1)^{2}}(\tilde{E}[x_{v,j}^{2}]\tilde{E}[q_{v,i}^{2}] + \tilde{E}[x_{v,i}^{2}]\tilde{E}[q_{v,j}^{2}]) \\
&= -\frac{k(k - 1)}{4} + \sum _{1 \leq i < j \leq k} \frac{k}{(k - 1)^{2}}(\tilde{E}[q_{v,i}^{2}] + \tilde{E}[q_{v,j}^{2}]) \\
&= -\frac{k(k - 1)}{4} + \binom{k}{2}\frac{k}{(k - 1)^{2}}2\tilde{E}[q_{v,1}^{2}] \\
&= -\frac{k(k - 1)}{4} + \frac{k^{2}}{k - 1}\tilde{E}[q_{v,1}^{2}].
\end{align*}
The value the SDP assigns is
\[ -\frac{(k - 1)^{2}}{4} + \sum _{i = 1} ^{k} \tilde{E}[q_{v,i}^{2}] = - \frac{(k - 1)^{2}}{4} + k\tilde{E}[q_{v,1}^{2}] \]
so the approximation ratio for a kinetic energy term is exactly $\frac{k}{k - 1}$. Thus the overall approximation ratio is
\[ \alpha_{k} \coloneqq \max\left(\alpha_{BOV,k}, \frac{k}{k - 1}\right). \]
By Proposition~\ref{prop:BOVconv}, we know $\alpha_{BOV, k} \to 1$ and thus $\alpha_{k} \to 1$ as $k \to \infty$. Thus for $k$ sufficiently large but constant, level 1 of this ncSoS relaxation beats the optimal product state approximation ratio.

\section{Acknowledgments}
The work in this paper was partially funded by a Doc Bedard fellowship from the Laboratory for Physical Sciences and MIT Center for Quantum Engineering. The author would like to thank Anand Natarajan and George Stepaniants for helpful discussions.

\nocite{*}
\printbibliography

\end{document}